\documentclass[12pt]{article}

\usepackage{amsmath,amssymb}
\usepackage{graphicx,color}

\newcommand{\amp}{&\!\!}
\newcommand{\mpl}{M_{\mbox{\tiny{Pl}}}}
\newcommand{\beq}{\begin{equation}}
\newcommand{\eeq}{\end{equation}}
\newcommand{\bea}{\begin{eqnarray}}
\newcommand{\eea}{\end{eqnarray}}
\newcommand{\lmfp}{\lambda_{\mbox{\tiny{MFP}}}}
\newcommand{\knl}{k_{\mbox{\tiny{NL}}}}
\newcommand{\lnl}{\lambda_{\mbox{\tiny{NL}}}}
\newcommand{\luv}{\lambda_{\mbox{\tiny{UV}}}}
\newcommand{\lbao}{\lambda_{\mbox{\tiny{BAO}}}}

\begin{document}

\title{{\bf The Effective Field Theory of Dark Matter and Structure Formation: Semi-Analytical Results}}

\author{
\Large{Mark P. Hertzberg} \\
~\\
{\em Stanford Institute for Theoretical Physics},\\
{\em Stanford University, Stanford, CA 94305, USA}\\
~\\
{\em Kavli Institute for Particle Astrophysics and Cosmology},\\
{\em Stanford University and SLAC, Menlo Park, CA 94025, USA}\\
~\\
{\em Center for Theoretical Physics, Dept.~of Physics},\\
{\em Massachusetts Institute of Technology, Cambridge, MA 02139, USA}}

\date{}

\maketitle 

\vspace{-0.8cm}

\begin{abstract}
Complimenting recent work on the effective field theory of cosmological large scale structures, here we present detailed approximate analytical results and further pedagogical understanding of the method. We start from the collisionless Boltzmann equation and integrate out short modes of a dark matter/dark energy dominated universe ($\Lambda$CDM)  whose matter is comprised of massive particles as used in cosmological simulations. This establishes a long distance effective fluid, valid for length scales larger than the non-linear scale $\sim$ 10\,Mpc, and provides the complete description of large scale structure formation. Extracting the time dependence, we derive recursion relations that encode the perturbative solution. This is exact for the matter dominated era and quite accurate in $\Lambda$CDM also. The effective fluid is characterized by physical parameters, including sound speed and viscosity. These two fluid parameters play a degenerate role with each other and lead to a relative correction from standard perturbation theory of the form $\sim 10^{-6}c^2\,k^2/H^2$. Starting from the linear theory, we calculate corrections to cosmological observables, such as the baryon-acoustic-oscillation peak, which we compute semi-analytically at one-loop order. Due to the non-zero fluid parameters, the predictions of the effective field theory agree with observation much more accurately than standard perturbation theory and we explain why. We also discuss corrections from treating dark matter as interacting or wave-like and other issues.
\let\thefootnote\relax\footnotetext{Electronic address: {\tt mphertz@stanford.edu, mphertz@mit.edu}}
\end{abstract}

\vspace*{-20cm} {\hfill MIT-CTP 4407\,\,\,\,} 



\newpage

\tableofcontents

\section{Introduction}

An effective field theory is a description of a  system that captures all the relevant degrees of freedom and describes all the relevant physics
at a macroscopic scale of interest. The short distance (``UV") physics is integrated out and affects the effective field theory only through various couplings in a perturbative expansion in the ratio of  microphysical UV scale/s to the macroscopic (``IR") scale being probed. 
This technique has been systematically used in particle physics and condensed matter physics for many years (e.g., see \cite{Georgi:1994qn,Manohar:1996cq,Kaplan:2005es}), but has not been fully used in astrophysics and cosmology.  The large scale properties of the universe acts as an important application and is in need of careful analysis. 

Of current fundamental  importance is to understand the initial conditions, contents, evolution and formation of the universe. It appears to be adequately described by the so-called $\Lambda$CDM cosmological model in which the matter content of the universe is primarily dark matter and the late time dark energy is adequately described by a cosmological constant. The early universe was dominated by a cosmic plasma in which baryons were tightly coupled to photons leading to so-called baryon-acoustic-oscillations.
The evidence for this model comes from a range of sources, including CMB data, lyman-$\alpha$ forrest, curvature constraints, supernovae type IA, weak lensing, and (of particular importance to the current discussion) structure formation, etc.
In this cosmological model, structure formation is primarily driven by the gravitational attraction of dark matter, which led to the clumping of baryons  including stars, galaxies, and clusters of galaxies. This arose from gravitational instabilities of the initial linear density fluctuations that were approximately adiabatic, scale-invariant, and Gaussian (e.g., see \cite{Peebles:2012mz,Peebles:2009hw,Primack:2006it,CervantesCota:2011pn}).

The power spectrum of large scale structure at late times is corrected from the initial linear input in interesting and important ways. For instance, non-linear effects alter the shape of the baryon-acoustic-oscillations in the power spectrum. The baryon-acoustic-oscillations are a vitally important probe of dark energy as they provide a standard ruler to constrain the cosmological expansion history (e.g., see \cite{Bassett:2009mm,Weinberg:2012es,Sanchez:2008iw,Wang:2006qt}). In general, one needs a proper understanding of the departures from linear theory in order to constrain fundamental physics, such as dark energy, primordial non-Gaussianity, and other probes of microscopic physics. 

There has been a substantial amount of work in the literature to understand  non-linear structure formation in the form of cosmological perturbation theory, including \cite{Bernardeau:2001qr,Jain:1993jh,Takahashi:2008yk,Shoji:2009gg,Jeong:2006xd,Crocce:2005xy,Crocce:2007dt,Matsubara:2007wj,McDonald:2006hf,Taruya:2007xy,Izumi:2007su,Matarrese:2007aj,Matarrese:2007wc,Nishimichi:2007xt,Peebles:2000yy,Carlson:2009it,Pietroni:2011iz,Tassev:2011ac,Roth:2011ru}. This includes what is usually referred to as ``standard-perturbation-theory" (SPT). In this approach, the continuity and Euler equations for a pressure-less and non-viscous dark matter fluid (with vanishing stress-tensor) is assumed. These non-linear equations for the dark matter are solved perturbatively around the linear solution and corrections are obtained order by order, usually truncated at the one-loop order. The theory involves integrating $k$-modes in the entire domain $0<k<\infty$, which involves treating all $k$-modes as perturbative. 

The analysis of the current paper stems from the fact that this ``standard" procedure necessarily has a qualitative and quantitative problem. The density fluctuations are not perturbative beyond a scale $\knl$, the ``non-linear scale"; the scale at which density fluctuations are $\delta\sim\mathcal{O}(1)$, where gravitational collapse may occur. 
This scale is roughly $\lambda_{NL}\sim10$\,Mpc or so.
Hence there are two regimes: $k<\knl$ which is weakly coupled and perturbative, and $k>\knl$ which is strongly coupled and non-perturbative. For cosmological purposes, we are normally interested in the low $k$-regime. However, these 2 regimes are coupled by non-linearities, so we must be very careful in attempts to describe the low $k$-regime perturbatively. The rigorous and complete method to do this is that of effective field theory. The procedure is to introduce some arbitrary cutoff $\Lambda$ on the $k$-modes of the fluid. We take this cutoff to be $\Lambda\lesssim \knl$, so that all modes of the fluid are perturbative. This means that the high $k$-modes ($k>\Lambda$) must be ``integrated out". In practice this means that these UV modes generate higher order derivative and non-linear corrections to the fluid equations for the low $k$-modes ($k<\Lambda$). We show that this includes terms such as pressure and viscosity; precisely the terms that are assumed to vanish in SPT. These terms are a real property of the dark matter fluid and they alter the power spectrum in an important and measurable way. These fluid parameters can be determined by matching to the full UV theory, i.e., N-body simulations. This furnishes an effective field theory for dark matter. This is a fluid that only involves weakly coupled modes and is connected to the full microphysical theory through these higher order operators. 

Important earlier work on this topic was performed in the very interesting Ref.~\cite{Baumann:2010tm}, where this basic conceptual foundation was laid out with particular focus on the issue of back-reaction at the scale of the horizon. In our recent work, which we continue here, we (i) focus on sub-horizon scales, (ii) obtain an explicit measurement of fluid parameters, (iii) perform an explicit computation of the power spectrum, and (iv) provide further insight and clarifications.
In the present paper we compliment and extend our important recent work in Ref.~\cite{Carrasco:2012cv} in which the measurement of fluid parameters was performed and the basic framework was put together.
Here we develop and describe in detail this effective fluid description of dark matter (and by extension; all matter, since baryons trace dark matter on large scales), which is the complete description of large scale structure formation. In particular, we recapitulate how to extract various fluid parameters from N-body simulations and then solve the effective fluid theory to some desired order in a perturbative expansion which we formulate recursively. 
We show how to approximately extract the time dependence in a $\Lambda$CDM universe, which connects in a simple and intuitive way with the previous standard perturbation theory, but highlights the essential differences arising from the fluid parameters. This leads to convenient and quite accurate results.
Our basic method and key results are summarized in the following discussion:

If we assume that the matter content of the universe is dominated by non-relativistic matter, primarily dark matter evolving under Newtonian gravity, we can smooth the corresponding collisionless Boltzmann equation in an expanding FRW background. This generates the usual continuity and Euler equations. An important point stressed in Refs.~\cite{Baumann:2010tm,Carrasco:2012cv} is that the latter includes an effective stress-tensor $[\tau^{ij}]_\Lambda$ that is sourced by the short-modes $\delta_s$.
By defining the effective stress-tensor by its correlation functions, it can be expanded in terms of the long density  $\delta_l$ and velocity $v_l^i$ fields as
\bea
[\tau^{ij}]_\Lambda\amp=\amp \delta^{ij}p_b+\rho_b\Bigg{[} c_s^2\,\delta^{ij}\delta_l-{c_{bv}^2\over Ha}\delta^{ij}\,\partial_k v_l^k\nonumber\\\amp-\amp{3\over4}{c_{sv}^2\over Ha}\left(\partial_jv_l^i+\partial_iv_l^j-{2\over3}\delta^{ij}\,\partial_kv_l^k\right)
\Bigg{]}+\ldots
\eea
The individual parameters $c_s^2$ and $c_v^2\equiv c_{sv}^2+c_{bv}^2$ are degenerate with each other at the one-loop order since we only track the growing modes, and degenerate with other parameters at higher loop order. While the shear viscosity parameter $c_{sv}^2$ affects the vorticity, which is a somewhat small effect.
As analyzed in Ref.~\cite{Carrasco:2012cv} one can directly evaluate the stress-tensor from the microphysical theory, i.e., from N-body simulations to extract such parameters. For  smoothing scale $\Lambda=1/3$\,[h/Mpc] at $z=0$ it is found
$c_s^2+f c_v^2\approx 9\times 10^{-7}c^2$, ($f$ is the logarithmic derivative of the growth function).
This direct measurement can also be obtained from matching to the power spectrum at some renormalization scale, resulting in a consistent value and a positive check on the validity of the theory.

We establish recursion relations for the density fluctuations and velocity field, allowing us to insert this measured value of the fluid parameter and obtain correlation functions. These parameters carry $\Lambda$ dependence which balances the $\Lambda$ dependence of the cutoff on the loops. If we take $\Lambda$ to large values the fluid parameters approach a finite quantity, representing the finite error made in the standard perturbation theory.
In particular the fluid parameters provide the following relative correction to the power spectrum (suppressing the time dependence and $\mathcal{O}(1)$ factors)
\beq
{\delta P(k)\over P_{L}(k)}\sim -{10^{-6}\,c^2\,k^2\over H^2}
\label{Pcorr}\eeq
where $P_{L}(k)$ is the linear power spectrum. 
Note that the pressure and viscosity act together to reduce the power spectrum by acting oppositely to gravity. 
This simple, but entirely real and rigorous correction to the power spectrum is essential to explain the observed shape of the baryon-acoustic-oscillations in the power spectrum relative to standard theory.

The outline of the paper is the following: In Section \ref{EffectiveFluid} we describe the basic theoretical setup. Operating in the Newtonian approximation in an expanding universe, we smooth the Boltzmann equation to obtain an effective fluid for cold dark matter. We recapitulate how to match its parameters to the microphysical results from N-body simulations. 
In Section \ref{PerturbationTheory} we solve the theory recursively for a matter dominated universe by extracting the time dependence, and lift this to $\Lambda$CDM in an approximate way also. This allows us to semi-analytically derive the power spectrum at one-loop order. 
In Section \ref{PowerNumResults} we present our numerical results for the power spectrum and compare to linear theory and standard perturbation theory.
In Section \ref{Discussion} we discuss the fluid's parameters, corrections from collisions, wave-like behavior, higher order moments, and the velocity field.
Finally, in Section \ref{Summary} we summarize the effective field theory and its role in  cosmology.

\section{Effective Fluid}\label{EffectiveFluid}

\subsection{Newtonian Approximation}

Cosmological perturbation theory around a flat FRW background may be performed in many gauges. One example is the Newtonian gauge. Scalar modes are captured by the following form for the metric
\beq
ds^2=-dt^2(c^2-2\phi({\bf x},t))+a(t)^2(1-2\psi({\bf x},t)/c^2)d{\bf x}^2
\eeq
were $a$ is the scale factor and ${\bf x}$ is a co-moving co-ordinate.
The Newtonian approximation is a valid  description for non-relativistic matter in an expanding background on sub-horizon scales, and will be sufficient for our purposes as we will study evolution of matter after equality. In this limit, only $\phi$ plays a role and not $\psi$.
Here $\phi$ is the Newtonian potential, sourced by  fluctuations in matter density \cite{Peebles}
\beq
\nabla^2\phi=4\pi G a^2(\rho({\bf x},t)-\rho_b(t))
\eeq
where $\rho$ is the matter density, which is a combination of dark matter and baryonic matter, and $\rho_b=\langle\rho\rangle$ is the background value, with $\rho_b(t)\propto1/a(t)^3$.
The Hubble parameter is determined by the Friedmann equation
\beq
H(t)^2={8\pi G\over3}(\rho_b(t)+\rho_{vac})
\eeq
in a flat $\Lambda$CDM universe. Here we allow for vacuum energy in $\rho_{vac}$, which we assume to be the cosmological constant of general relativity, as is consistent with all current data.
The $\Lambda$CDM concordance model indicates that this is a valid description of the universe for all times well after matter-radiation equality. Furthermore, after the baryons decouple from the photons,  the baryons tend to trace the dark matter on large scales, leading to a single non-relativistic fluid that we will describe.

\subsection{Phase Space Evolution}

We treat dark matter as classical point particles and ignore its quantum nature. This is a very good approximation for most dark matter candidates, but can breakdown for extremely light axions which organize into a state of very high occupation number, with quantum pressure and associated sound speed $c_s\sim {\hbar\, k\over a\,m_a}$ \cite{Sikivie:2009qn}. For QCD axions, this is ignorable on large scales (since it vanishes for small $k$) and will be ignored here; see Section \ref{Wave} for further discussion.
Indeed N-body simulations are ordinarily done with a set of classical point particles. At each moment in time the output is a set of N-vectors which we label $n$,
with co-moving co-ordinates ${\bf x}_n$, and proper peculiar velocity ${\bf v}_n$.

Let $f_{n}({\bf x},{\bf p})$ be the single particle phase space density defined such that $f_{n}({\bf x},{\bf p})\,d^3{\bf x}\,d^3{\bf p}$ is the probability of particle $n$ occupying an infinitesimal phase space element. 
For a point particle, the phase space density is
\beq
f_n({\bf x},{\bf p})=\delta^3_D({\bf x}-{\bf x}_n)\,\delta^3_D({\bf p}-m\,a\,{\bf v}_n)
\eeq
(where both ${\bf x}$ and ${\bf p}$ are co-moving).
By summing over $n$, we define the total phase space density $f$, mass density $\rho$, momentum density $\pi^i$, kinetic-tensor $\sigma^{ij}$ as
\bea
f({\bf x},{\bf p})\amp=\amp \sum_n \delta^3_D({\bf x}-{\bf x}_n)\,\delta^3_D({\bf p}-m\,a\,{\bf v}_n)\\
\rho({\bf x})\amp=\amp m\,a^{-3}\!\int d^3{\bf p}\,f({\bf x},{\bf p})\nonumber\\ \amp=\amp\sum_nm\,a^{-3}\,\delta^3_D({\bf x}-{\bf x}_n)\\
\pi^i({\bf x})\amp=\amp a^{-4}\!\int d^3{\bf p}\,p^if({\bf x},{\bf p})\nonumber\\ \amp=\amp\sum_nm\,a^{-3}\,v_n^i\,\delta^3_D({\bf x}-{\bf x}_n) \\
\sigma^{ij}({\bf x})\amp=\amp m^{-1}a^{-5}\!\int d^3{\bf p}\,p^i\,p^jf({\bf x},{\bf p})\nonumber\\ \amp=\amp\sum_nm\,a^{-3}\,v_n^i\,v_n^j\,\delta^3_D({\bf x}-{\bf x}_n) 
\eea

The Newtonian potential is sensitive to an infrared quadratic divergence in an infinite homogeneous universe.
To isolate this divergence we introduce an exponential infrared regulator with cutoff $\mu$ (a ``mass" term) and will take the $\mu\to 0$ limit whenever it is allowed. 
The Newtonian potential $\phi$ is
\bea
\phi_n({\bf x})\amp=\amp-G\,a^2\!\int d^3{\bf x}'{\rho_n({\bf x}')\over|{\bf x}-{\bf x}'|}e^{-\mu|{\bf x}-{\bf x}'|}\nonumber\\\amp=\amp-{mG\over a|{\bf x}-{\bf x}_n|}e^{-\mu|{\bf x}-{\bf x}_n|}\\
\phi({\bf x})\amp=\amp-G\,a^2\!\int d^3{\bf x}'{\rho({\bf x}')-\rho_b\over|{\bf x}-{\bf x}'|}e^{-\mu|{\bf x}-{\bf x}'|}\nonumber\\\amp=\amp\sum_n\phi_n
+{4\pi Ga^2\rho_b\over\mu^2}
\eea
Note that 
\beq
\mu^2\sum_n\phi_n\to-4\pi G a^2\rho_b\,\,\,\,\,\,\mbox{as}\,\,\,\,\,\, \mu\to 0
\eeq
The $k$-space version of the Newtonian potential is 
\bea
\phi_n({\bf k})\amp=\amp-{4\pi mG\over a(k^2+\mu^2)}e^{-i{\bf k}\cdot{\bf x}_n}\\
\phi({\bf k})\amp=\amp\sum_n\phi_n({\bf k})+{4\pi G a^2\rho_b\over \mu^2}(2\pi)^3\delta^3_D({\bf k})
\eea
where the final term evidently subtracts out the zero-mode.

\subsubsection{Boltzmann equation}
Cold dark matter candidates, such as WIMPs and axions, have very small scattering cross sections with itself and standard model particles.
Here we make the approximation that we can ignore the scattering altogether (see Section \ref{Collisions} for further discussion). 
Restricting our attention then to collisionless classical particles interacting only via gravity, the Boltzmann equation is
\beq
\left(p^{\mu}{\partial\over\partial x^{\mu}}+\Gamma^{\mu}_{\alpha\beta}p^{\alpha}p^{\beta}{\partial\over\partial p^{\mu}}\right)f_n=0
\eeq
In the Newtonian limit  in a flat FRW expanding background, the collisionless Boltzmann equation becomes
\beq
0={Df_{n}\over Dt}={\partial f_{n}\over\partial t}+{{\bf p}\over m\,a^2}\cdot{\partial f_{n}\over\partial{\bf x}}-m\!\left(\sum_{\bar{n}\neq n}{\partial\phi_{\bar{n}}\over\partial{\bf x}}\right)\!\cdot{\partial f_{n}\over\partial{\bf p}}
\eeq
where the self-force has been subtracted out of the sum over $\bar{n}$ in the last term.
By summing over $n$ we have
\beq
0={Df\over Dt}={\partial f\over\partial t}+{{\bf p}\over m\,a^2}\cdot{\partial f\over\partial{\bf x}}-m\sum_{\bar{n}\neq n}{\partial\phi_{\bar{n}}\over\partial{\bf x}}\cdot{\partial f_{n}\over\partial{\bf p}}\label{Boltzmann}
\eeq
where the final term now involves a double summation over $\bar{n}$ and $n$.

\subsection{Smoothing}

From this we would like to construct various quantities that describe an effective fluid at large length scales \cite{Baumann:2010tm}.
Most importantly, we need the effective stress-tensor $\tau^{ij}$ that is sourced by the short wavelength modes. In order to define a fluid we must perform a smoothing
of the output data from the simulation. 

To this end let us define the following Gaussian smoothing function
\beq
W_{\Lambda}({\bf x})\equiv \left({\Lambda\over\sqrt{2\pi}}\right)^{\!3}\exp\left(-{\Lambda^2{\bf x}^2\over2}\right)
\eeq
(here $\Lambda$ is a smoothing scale and should not be confused with the cosmological constant) which is normalized such that $\int d^3{\bf x}\,W({\bf x})=1$. Of course our results will not depend on the choice of smoothing function, but the Gaussian is chosen for convenience. (In fact we will later include numerical results for the sinc function corresponding to a step function in $k$-space).
In $k$-space the Gaussian smoothing function is 
\beq
W_\Lambda(k)=\exp\left(-{k^2\over2\,\Lambda^2}\right)
\eeq
This will smooth the fluid quantities on (co-moving) length scales $\gg\Lambda^{-1}$, acting as a cutoff on modes $k\gtrsim\Lambda$. We should choose $\Lambda\lesssim \knl$, were $\knl$ is the wavenumber where modes have become non-linear, so that we integrate out the non-linear modes. Given the form of $W_\Lambda$, we can estimate a rough value for the smoothing scale in position space as $\luv\sim 2\pi/\Lambda$.

For certain observables $O({\bf x})$, we will define the smoothed value by the convolution
\beq
O_l({\bf x})=[O]_\Lambda({\bf x})\equiv\int d^3{\bf x}'\,W_\Lambda({\bf x}-{\bf x}')O({\bf x}')
\eeq

The smoothed versions of the phase space density $f_l$, mass density $\rho_l$, momentum density $\pi^i_l$, stress-tensor $\sigma^{ij}_l$, 
derivative of Newtonian potential $\partial_i\phi_l$ are
\bea
f_l({\bf x},{\bf p})\amp=\amp\sum_n W_\Lambda({\bf x}-{\bf x}_n)\,\delta^3_D({\bf p}-m\,a\,{\bf v}_n)\\
\rho_l({\bf x})\amp=\amp\sum_nm\,a^{-3}\,W_\Lambda({\bf x}-{\bf x}_n)\\
\pi_l^i({\bf x})\amp=\amp\sum_nm\,a^{-3}\,v^i_n\,W_\Lambda({\bf x}-{\bf x}_n)\\
\sigma^{ij}_l({\bf x})\amp=\amp\sum_nm\,a^{-3}\,v^i_n\,v^j_n\,W_\Lambda({\bf x}-{\bf x}_n)
\eea
The ``l" substrict indicates that these only depend on the {\em long} modes.
Similarly, the smoothed version of the Newtonian potential $\phi_l$ is
\bea
\phi_{l,n}({\bf x})\amp=\amp-{mG\over a|{\bf x}-{\bf x}_n|}\mbox{Erf}\!\left[{\Lambda|{\bf x}-{\bf x}_n|\over\sqrt{2}}\right]e^{-\mu|{\bf x}-{\bf x}_n|}\,\,\\
\phi_l({\bf x})\amp=\amp\sum_n\phi_{l,n}+{4\pi Ga^2\rho_b\over\mu^2}
\eea
Here we used the Gaussian smoothing function to explicitly evaluate $\phi_l({\bf x})$ in terms of the error function
$\mbox{Erf}(z)\equiv{2\over\sqrt{\pi}}\int_0^z dt\,e^{-t^2}$.

We now write down the smoothed version of (\ref{Boltzmann}) by multiplying it by $W_\Lambda$ and integrating over space 
\bea
0\amp=\amp\left[{Df\over Dt}\right]_\Lambda={\partial f_l\over\partial t}+{{\bf p}\over m\,a^2}\cdot{\partial f_l\over\partial{\bf x}}\nonumber\\
\amp-\amp m\sum_{n\neq\bar{n}}\bar\int d^3{\bf x}'\,W_\Lambda({\bf x}-{\bf x'}){\partial\phi_n\over\partial{\bf x}'}({\bf x}')\cdot{\partial f_{\bar{n}}\over\partial{\bf p}}({\bf x}',{\bf p})\,\,\,\,\,\,\,\,\,
\eea
where we used linearity to express the first two terms directly in terms of $f_l$. However the third term is more complicated.
The equations of motion are obtained by taking moments of this smoothed Boltzmann equation, i.e.,
\beq
0=\int d^3{\bf p}\,p^{i_1}\cdots p^{i_m}\left[{Df\over Dt}\right]_\Lambda\!\!({\bf x},{\bf p})
\eeq
One often subtracts out traces of the tensor structure of the higher order moments for convenience. Here we will only make use of the zeroth and first moments, so this detail is not important for us.

\subsection{Effective Continuity Equation}

The zeroth moment gives the continuity equation, which we find to be
\beq
\dot{\rho}_l+3\,H\rho_l+{1\over a}\partial_i(\rho_l \, v_l^i)=0
\eeq
Here we introduced the velocity field 
\beq
v_l^i({\bf x})\equiv{\pi_l^i({\bf x})\over\rho_l({\bf x})}=
{\sum_n v^i_n\,W_\Lambda({\bf x}-{\bf x}_n)\over \sum_n W_\Lambda({\bf x}-{\bf x}_n)}.
\eeq
The continuity equation relates the zeroth moment of the phase space distribution, $\rho_l$, to the first moment of the phase space distribution, $\rho_l v_l^i$.
We now turn to the next moment of the Boltzmann equation to obtain an equation for the velocity field itself.

\subsection{Effective Euler Equation}

The first moment gives the Euler equation, which we find to be
\beq
\dot{v}_l^i+H v_l^i+{1\over a}v_l^j\partial_j v_l^i+{1\over a}\partial_i\phi_l=-{1\over a\,\rho_l}\partial_j\!\left[\tau^{ij}\right]_\Lambda
\eeq
where the effective stress-tensor that is sourced by the short modes is given by
\beq
[\tau^{ij}]_\Lambda=\kappa_l^{ij}+\Phi_l^{ij}\label{stressT}
\eeq 
The Euler equation relates the first moment of the phase space distribution $v_l^i$, to the second moment of the phase space distribution, $\sigma_l^{ij}$.
Here $\kappa^{ij}_l$ is a type of kinetic dispersion and $\Phi^{ij}_l$ is a type of gravitational dispersion, namely
\bea 
\kappa^{ij}_l\amp=\amp\sigma_l^{ij}-\rho_lv_l^iv_l^j\label{kappal}\label{kinT}\\
\Phi^{ij}_l\amp =\amp-{w^{kk}_l\delta^{ij}-2w^{ij}_l\over 8\pi G\,a^2}+{\partial_k\phi_l\partial_k\phi_l\delta^{ij}-2\partial_i\phi_l\partial_j\phi_l\over 8\pi G\,a^2}\label{gravT}\,\,\,\,\,\,\,\,
\eea
where
\bea
w^{ij}_l({\bf x}) =
\int d^3{\bf x}'\,W_\Lambda({\bf x}-{\bf x}')\!\amp\!\!\Big{[}\amp\!\!\!\partial_{i'}\phi({\bf x}')\,\partial_{j'}\phi({\bf x}')\nonumber\\
-\sum_n \amp\amp \!\!\!\!\!\!\partial_{i'}\phi_n({\bf x}')\,\partial_{j'}\phi_n({\bf x}') \Big{]}\,\,\,\,\label{weqn}
\eea
Note that we have subtracted out the self term in $w^{ij}_l$,
and used $\nabla^2\phi=4\pi Ga^2(\rho-\rho_b)$ and $\nabla^2\phi_l=4\pi Ga^2(\rho_l-\rho_b)$ to express $\Phi_l$ in terms of $\phi$ and $\phi_l$.
In the limit in which there are no short modes, it is simple to see from the definition of $\kappa_l$ and $\Phi_l$ that they vanish in this limit. More on the mathematical details of this are given in Appendix \ref{Shortmodes}.

\subsubsection{Derivative of Stress-Tensor}

In the general relativistic theory the absolute value of the stress-tensor will act as a source for gravity, and could, in principle, be important at the scale of the horizon. This includes the trace of the stress-tensor that is a type of pressure, which we discuss in Appendix \ref{TraceTensor}.
On sub-horizon scales, however, a non-relativistic analysis is applicable in which the right hand side of the Euler equation only involves the derivative of the stress-tensor:
\beq
J_l^i={1\over a\,\rho_b}\partial_j[\tau^{ij}]_\Lambda
\eeq
here we divided by the background density $\rho_b$ for convenience.
This quantity will be quite important in out analysis, and takes on the following explicit form
\bea
a\,\rho_b\, J_l^i=\partial_j(\sigma_l^{ij}-\rho_lv_l^iv_l^j)+\sum_{n\neq\bar{n}}\left[\rho_n\,\partial_i\phi_{\bar{n}}\right]_\Lambda-\rho_l\partial_i\phi_l\,\,\,\,
\eea
The smoothed quadratic form can be expressed as
\bea
\left[\rho_n\,\partial_i\phi_{\bar{n}}\right]_\Lambda={m^2G(x_n^i-x_{\bar{n}}^i)\over a^4|{\bf x}_n-{\bf x}_{\bar{n}}|^3}\!\amp(\amp \!\! 1+\mu|{\bf x}_n-{\bf x}_{\bar{n}}|)
e^{-\mu|{\bf x}_n-{\bf x}_{\bar{n}}|}W_\Lambda({\bf x}-{\bf x}_n)\,\,\,\,\,
\eea
which requires one to perform a double summation over $n$ and $\bar{n}$, which can be computationally expensive.
Notice, however, that while the stress-tensor involves an integral over the gravitational potential in eq.~(\ref{weqn}), the derivative of the stress-tensor does not require this (by making use of the Poisson equation and then integrating over the delta-functions).

\subsection{Fluid Parameters}

\begin{figure}[t]
\center{\includegraphics[width=10cm]{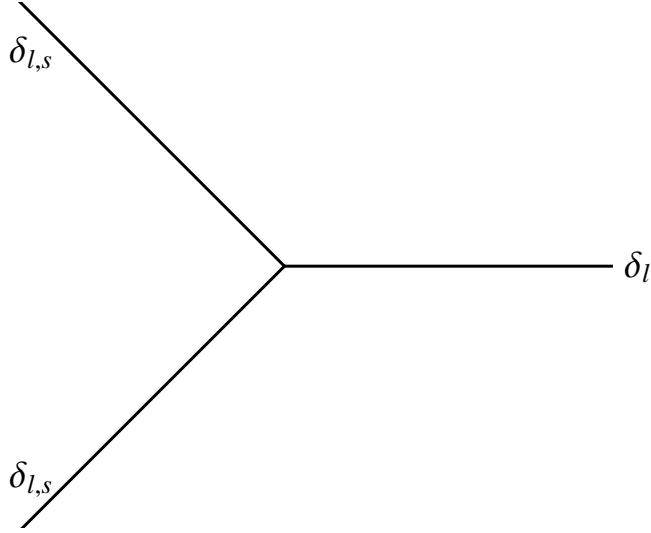}}
\caption{Vertex for interaction between long-long mode coupling or long-short mode coupling.}
\label{FeynmanVertex}\end{figure}

The effective stress-tensor $[\tau^{ij}]_\Lambda$ comes from smoothing over the short modes and therefore is ``sourced" by the short modes.
However, the important quantities that will arise when we later compute $N$-point functions, such as the 2-point function, involves correlation functions of the stress-tensor with the long modes,
\beq
v_l^i({\bf x})\,\,\,\,\,\,\mbox{and}\,\,\,\,\,\,\delta_l({\bf x})\equiv{\rho_l({\bf x})\over\rho_b}-1
\eeq 
(note that $\phi_l$ is determined as a constrained variable through the Poisson equation $\nabla^2\phi_l=4\pi Ga^2\rho_b\,\delta_l$).
For instance, the expectation value $\langle[\tau^{ij}]_\Lambda\rangle$ is some background pressure. More importantly though is the mode-mode coupling. Coupling between long modes is connected to the non-linear terms in the continuity and Euler equations. While coupling between long and short modes is connected to the stress-tensor, which generates non-zero correlation functions $\langle[\tau^{ij}]_\Lambda\,\delta_l\rangle$ and $\langle[\tau^{ij}]_\Lambda\,v_l^k\rangle$. The long-long and short-short mode couplings are represented by vertices in Fig.~\ref{FeynmanVertex}.
Note that the dark matter gas is non-thermal, and this indicates that the stress-tensor cannot be derived by some analytical thermal argument, such as would be the case for air under ordinary conditions. However, the reason one can make progress is to recognize that the stress-tensor organizes itself in terms of length scales with coefficients that come from matching; this organization is guaranteed by the principles of effective field theory.

In order to extract this dependence of the stress-tensor on the long modes, we implicitly write the stress-tensor as an expansion in terms of the long fields, whose coefficients are determined by various correlation functions. This will involves a type of pressure perturbation term $\propto \delta^{ij}\, \delta_l$, 
a shear viscosity term $\propto \partial^jv_l^i+\partial^iv^j_l-{2\over3}\delta^{ij}\,\partial_kv_l^k$, and a bulk viscosity term $\propto \delta^{ij}\,\partial_k v_l^k$.
Demanding rotational symmetry, we write a type of effective field theory expansion for the stress-tensor as
\bea
[\tau^{ij}]_\Lambda\amp=\amp\rho_b\Bigg{[} c_s^2\,\delta^{ij}(\gamma^{-1}+\delta_l)-{c_{bv}^2\over Ha}\delta^{ij}\,\partial_k v_l^k\nonumber\\\amp-\amp{3\over4}{c_{sv}^2\over Ha}\left(\partial^jv_l^i+\partial^iv_l^j-{2\over3}\delta^{ij}\,\partial_kv_l^k\right)
\Bigg{]}+\Delta\tau^{ij}\,\,\,\,\,\,\,\,\,\,\,\,\label{Ansatz}
\eea
where $\gamma$ would correspond to the ratio of specific heats in an ordinary fluid (e.g., $\gamma=5/3$ for an ideal monatonic gas) but here it just parameterizes the background pressure term, $c_s$ is a sound speed, and $c_{sv}\,,c_{bv}$ are viscosity coefficients with units of speed. Note that $c_s,\,c_{sv},\,c_{bv}$ are allowed to depend on time, but not space. Our fluid coefficients are related to the conventional fluid quantities: background pressure $p_b$, pressure perturbation $\delta p$, shear viscosity $\eta$, and bulk viscosity $\zeta$ by
\bea
&& p_b={c_s^2\rho_b\over\gamma},\,\,\,\,\,\,\,\delta p=c_s^2\rho_b\delta_l,\nonumber\\
&& \eta={3\rho_b c_{sv}^2\over 4H},\,\,\,\,\,\,\,\zeta={\rho_b c_{bv}^2\over H}
\eea
In addition to what is included in (\ref{Ansatz}) there is an entire tower of higher order corrections carrying the appropriate rotational symmetry, guaranteed to exist by the principles of effective field theory. 
These will be parametrically suppressed at low wave number $k$ compared to the non-linear wavenumber $k_{NL}$, and will not enter to the order we shall work (which will be $\mathcal{O}(\delta^4)$), and so shall be ignored in the present discussion. Here $\Delta\tau^{ij}$ represents stochastic fluctuations due to fluctuations in the short modes, with $\langle\Delta\tau^{ij}\rangle=0$. We will return to this issue  in Section \ref{StochasticFluctuations}.

For convenience, let us define the following quantities from the stress-tensor
\bea
J_l^i\amp\equiv\amp{1\over a\,\rho_b}\partial_j\!\left[\tau^{ij}\right]_\Lambda\\
A_l^{ki}\amp\equiv\amp{1\over a^2\rho_b}\partial_k\partial_j\!\left[\tau^{ij}\right]_\Lambda=\partial_kJ_l^i/a\\
A_l\amp\equiv\amp{1\over a^2\rho_b}\partial_i\partial_j\!\left[\tau^{ij}\right]_\Lambda=\partial_i J_l^i/a\\
B_l\amp\equiv\amp{1\over a^2\rho_b}\left(\partial_i\partial_j-{\delta^{ij}\over3}\partial^2\right)\!\left[\tau^{ij}\right]_\Lambda
\eea
Now we introduce the dimensionless velocity divergence
\beq
\Theta_l\equiv-{\partial_kv_l^k\over Ha},\,\,\,\,\,\,\,\,\,\,\,\label{veldiv}
\Theta^{ki}_l\equiv-{\partial_kv_l^i\over Ha}
\eeq
Then according to our ansatz (\ref{Ansatz}) (and ignoring stochastic fluctuations for now) we have
\bea
a\,J_l^i\amp=\amp c_s^2\,\partial_i\delta_l+{3\over 4}c_{sv}^2\,\partial_j\Theta_l^{ji}
+\left({c_{sv}^2\over4}+c_{bv}^2\right)\partial_i\Theta_l\\
a^2A_l^{ki}\amp=\amp c_s^2\,\partial_k\partial_i\delta_l+{3\over 4}c_{sv}^2\,\partial_k\partial_j\Theta_l^{ji}
+\left({c_{sv}^2\over4}+c_{bv}^2\right)\partial_k\partial_i\Theta_l\label{Alki}\\
a^2A_l\amp=\amp c_s^2\,\partial^2\delta_l+(c_{sv}^2+c_{bv}^2)\partial^2\Theta_l\label{Al}\\
a^2B_l\amp=\amp c_{sv}^2\,\partial^2\Theta_l\label{Bl},
\eea

In order to extract the coefficients, we multiply each of these by the functions on the right hand side and then 
form a position space correlation function $\langle\ldots\rangle$, say $\left\langle\psi_1({\bf x}+{\bf x}')\psi_2({\bf x}')\right\rangle$. 
We will need the following set of correlation functions 
\bea
P_{A\,\delta}(x)\amp\equiv\amp\left\langle A_l({\bf x}+{\bf x}')\,\delta_l({\bf x}')\right\rangle\\
P_{A\,\Theta}(x)\amp\equiv\amp\left\langle A_l({\bf x}+{\bf x}')\,\Theta_l({\bf x}')\right\rangle\\
P_{A^{ki}\,\Theta^{ki}}(x)\amp\equiv\amp\left\langle A_l^{ki}({\bf x}+{\bf x}')\,\Theta_l^{ki}({\bf x}')\right\rangle\\
P_{B\,\Theta}(x)\amp\equiv\amp\left\langle B_l({\bf x}+{\bf x}')\,\Theta_l({\bf x}')\right\rangle\\
P_{\delta\,\delta}(x)\amp\equiv\amp\left\langle\delta_l({\bf x}+{\bf x}')\,\delta_l({\bf x}')\right\rangle\\
P_{\delta\,\Theta}(x)\amp\equiv\amp\left\langle\delta_l({\bf x}+{\bf x}')\,\Theta_l({\bf x}')\right\rangle\\
P_{\Theta\,\Theta}(x)\amp\equiv\amp\left\langle\Theta_l({\bf x}+{\bf x}')\,\Theta_l({\bf x}')\right\rangle\\
P_{\Theta^{ji}\,\Theta^{ki}}(x)\amp\equiv\amp\left\langle\Theta_l^{ji}({\bf x}+{\bf x}')\,\Theta_l^{ki}({\bf x}')\right\rangle
\eea
By rearranging, we find the following expressions for the fluid parameters
\bea
c_s^2\amp=\amp{P_{A\,\Theta}(x)\,\partial^2P_{\delta\,\Theta}(x)-P_{A\,\delta}(x)\,\partial^2P_{\Theta\,\Theta}(x)
\over (\partial^2 P_{\delta\,\Theta}(x))^2/a^2-\partial^2P_{\delta\,\delta}(x)\,\partial^2P_{\Theta\,\Theta}(x)/a^2}\label{csx}\\
c_{v}^2\amp=\amp {P_{A\,\delta}(x)\,\partial^2P_{\delta\,\Theta}(x)-P_{A\,\Theta}(x)\,\partial^2P_{\delta\,\delta}(x)
\over (\partial^2P_{\delta\,\Theta}(x))^2/a^2-\partial^2P_{\delta\,\delta}(x)\,\partial^2P_{\Theta\,\Theta}(x)/a^2} \label{cvisx} \,\,\,\,\,\,  \\  
c_{sv}^2\amp=\amp{4\over3}{P_{A^{ki}\,\Theta^{ki}}(x) - P_{A\,\Theta}(x)\over \partial^2 P_{\Theta^{ki}\,\Theta^{ki}}(x)/a^2 -\partial^2 P_{\Theta\,\Theta}(x)/a^2}
\,\,\,\,\,\,\,\mbox{}\,\,\nonumber\\
 \amp=\amp {P_{B\,\Theta}(x)\over\partial^2P_{\Theta\,\Theta}(x)/a^2}
\label{csvx}
\eea
where 
\beq
c_{v}^2\equiv c_{sv}^2+c_{bv}^2
\eeq
 is the sum of the viscosity coefficients.
The final result for each of the fluid coefficients will surely have some spatial dependence, 
so one should take the $x\gg\lnl$ limit of the final result to extract a constant value. 
Note that in eq.~(\ref{csvx}) we have provided two alternate expressions for $c_{sv}^2$.
In the linear theory, $\delta\sim\Theta$, allowing one to approximate the sum 
\beq
c_s^2+c_v^2\approx {P_{A\Theta}(x)\over\partial^2P_{\Theta\Theta}(x)/a^2}
\approx {P_{A\delta}(x)\over\partial^2P_{\delta\delta}(x)/a^2}.
\eeq

These expressions may also be given in $k$-space by taking the $k\ll \knl$ limit. To do so we Fourier transform each of the correlation functions.
We define the Fourier transform as
\beq
O({\bf k})\equiv\int d^3{\bf x}\,e^{-i{\bf k}\cdot{\bf x}}O({\bf x})
\eeq
By translational invariance, each $k$-space correlation functions takes the form
\beq
\left\langle\psi_1({\bf k})\,\psi_2({\bf k}')\right\rangle=(2\pi)^3\delta^3_D({\bf k}+{\bf k}')\,P_{\psi_1\,\psi_2}(k)
\eeq
where $P_{\psi_1\,\psi_2}(k)$ is the Fourier transform of $P_{\psi_1\,\psi_2}(x)$. 
This allows us to write the fluid parameters as
\bea
c_s^2\amp=\amp{P_{A\,\Theta}(k)\,P_{\delta\,\Theta}(k)-P_{A\,\delta}(k)\,P_{\Theta\,\Theta}(k)
\over -k^2 P_{\delta\,\Theta}(k)^2/a^2+k^2 P_{\delta\,\delta}(k)\,P_{\Theta\,\Theta}(k)/a^2}\label{csk}\\
c_{v}^2\amp=\amp {P_{A\,\delta}(k)\,P_{\delta\,\Theta}(k)-P_{A\,\Theta}(k)\,P_{\delta\,\delta}(k)
\over -k^2 P_{\delta\,\Theta}(k)^2/a^2+k^2 P_{\delta\,\delta}(k)\,P_{\Theta\,\Theta}(k)/a^2} \label{cvisk}\\
c_{sv}^2\amp=\amp{4\over3}{P_{A^{ki}\,\Theta^{ki}}(k) - P_{A\,\Theta}(k)\over -k^2 P_{\Theta^{ki}\,\Theta^{ki}}(k)/a^2 +k^2 P_{\Theta\,\Theta}(k)/a^2}
\,\,\,\,\,\,\,\mbox{}\,\,\nonumber\\
 \amp=\amp {P_{B\,\Theta}(k)\over-k^2P_{\Theta\,\Theta}(k)/a^2}\label{csvk}
\eea
We shall soon who that the combination $c_s^2+c_v^2$ is most important for the leading correction to the power spectrum.
Whether it is possible to measure each of these parameters individually, or only certain combinations, is discussed later in Section \ref{Degeneracy}.

\subsection{Relative Size of Terms}

Let us recall the primordial power spectra. In the first part of this subsection we will focus on a pure Einstein de Sitter universe, i.e., ignore the turn-over in the power spectrum due to the transfer function $T(k)$ (see next Section for its description). We will then include comments on the change that occurs when the transfer function is included, and all our numerical results in the latter part of this paper will be for the real universe including the full transfer function.
 
The primordial power spectrum in the Newtonian potential is approximately scale invariant (in this section we will suppress factors of $2\pi$)
\beq
P_{\phi\phi}(k)\sim {10^{-10}c^4\over k^3}
\eeq

For sub-horizon modes that entered the horizon in a matter dominated era, the Poisson equation gives $k^2\phi_k=-{3\over2}H^2a^2\,\delta_k$ 
and the corresponding power spectra is
\beq
P_{\delta\delta}(k)\sim{10^{-10}c^4k\over H^4 a^4}
\eeq
This means that the characteristic value of $\delta_l$ in position space, on a scale set by $k$, is
\beq
\delta_l\sim\sqrt{k^3\,P_{\delta\delta}(k)}\sim 10^{-5}\left({ck\over Ha}\right)^2
\eeq
This estimate is valid for $k\lesssim k_{eq}$. For $k>k_{eq}$ the rise in $\delta_l$ is only logarithmic. 

Let us now compare the relative size of the terms that appear in the Euler equation. 
We use the Hubble friction term $H v_l^i$ as the quantity to compare to. 
The relative size of the pressure $c_s^2\,\partial_i\delta_l/a$ or viscosity $c_v^2\,\partial_i\partial_k v_l^k/(Ha^2)\sim c_v^2\,\partial_i\delta_l/a$ is 
\bea
{\mbox{Pressure,Viscosity}\over\mbox{Hubble Friction}}\amp\sim\amp{c_{s,v}^2\, k \,\delta_l/a\over Hv_l}\\
\amp\sim\amp c_{s,v}^2\left({k\over Ha}\right)^2\sim{c_{s,v}^2\over 10^{-5}c^2}\,\delta_l
\eea
The relative size of the non-linear piece of the velocity convective derivative $v_l^i\partial_jv_l^i/a$ is
\bea
{\mbox{non-linear Velocity}\over\mbox{Hubble Friction}}\amp\sim\amp{k\,v_l^2/a\over Hv_l}\\
\amp\sim\amp 10^{-5}\left({ck\over Ha}\right)^2\sim\delta_l
\eea
Note that since we expect $c_s^2\sim c_v^2\sim 10^{-5}c^2$ for a pure Einstein de Sitter universe, then the pressure, viscosity, and non-linear velocity piece appear to be comparable on all $k$-scales
and all are $\sim\delta_l$. This also means that the terms $\delta^{(1)}J^{i\,(1)}$ and $J^{i\,(2)}$, which appear in the Taylor expanded Euler equation, are suppressed by a factor of $\sim\delta_l^2$ and so are all higher order again.
These estimates are consistent with the fact that we should only probe scales larger than the non-linear scale, i.e., $k^{-1}\gtrsim \knl^{-1}\sim 10^{-2.5}c/(Ha)$.

On the other hand, our analysis is not applicable in the regime approaching the horizon scale where general relativistic (GR) corrections will be important.
For instance, GR will change the partial derivatives to covariant derivatives leading to Hubble corrections. This leads us to estimate
\beq
{\mbox{GR Correction}\over\mbox{Newtonian Approximation}}\sim\left({Ha\over c\,k}\right)^2\sim{10^{-5}\over\delta_l}
\eeq
In other words, we should only probe scales smaller than Hubble, i.e., $k^{-1}\lesssim k_H^{-1}\sim c/(Ha)$.

For baryon-acoustic-oscillations the relevant scale is roughly $k_{dec}^{-1}\sim{c\over\sqrt{3}}(H_{dec}\,a_{dec})^{-1}$ at the time of decoupling,
where ${c\over\sqrt{3}}$ is the sound speed of the photon-baryon plasma (the associated peak in the fluctuations will be at a somewhat smaller scale when all modes are properly included). Red-shifting to today gives an estimate for $\delta_l$ that is somewhat less than 1 on this scale.
So this appears to fit nicely in the window where our approximations are valid. However, this estimate is overly simplistic, as it ignores the turn-over in the power spectrum for modes that enter during the radiation dominated era as described by the transfer function $T(k)$. Nevertheless, this qualitatively sets the basic hierarchy.

Turning then to the various length scales in the real universe: the characteristic length scale for baryon-acoustic-oscillations is $\lbao\sim 120$\,Mpc (see Fig.~\ref{CAMB2}). The non-linear scale is not sharply defined, since the power spectrum turns over to a logarithm for modes that entered in the radiation dominated era, but a characteristic value is $\lnl\sim 10$\,Mpc. This suggests we put the cutoff scale on the effective fluid at $\luv\sim2\pi/\Lambda\sim20$\,Mpc, or so, in order for it to be just above the non-linear  scale. The Hubble scale $d_H\sim 4$\,Gpc is a lot larger and so too is the scale of equality $\lambda_{eq}\sim 600$\,Mpc. A summary of this hierarchy is
\beq
\lnl<\luv<\lbao<\lambda_{eq}<d_{H}
\eeq
Although one should be careful here; although the baryon-acoustic-oscillation scale is $\sim 120$\,Mpc, its width is much smaller $\sim 20$\,Mpc. So one may actually need a somewhat smaller $\luv$ than $\sim 20$\,Mpc to fully resolve the baryon-acoustic-oscillation peak. 
But this then starts to push up against the non-linear scale $\sim 10$\,Mpc. However, we will eventually send $\luv\to 0$, once we have removed the cutoff dependence, so this is not necessarily a problem.

\section{Perturbation Theory}\label{PerturbationTheory}

\subsection{Linear Power Spectrum}

Inflation generates the primordial power spectrum, which we assume to be Gaussian
\beq
\langle\delta_L({\bf k})\delta_L({\bf k}')\rangle=(2\pi)^3\delta_D^3({\bf k}+{\bf k}')P_{\mbox{\tiny{inf}}}(k)
\eeq
where the ``L" subscript indicates that the initial fluctuations are in the linear regime.
Here $P_{\mbox{\tiny{inf}}}\propto k^{n_s}$, with $n_s=1$ for a scale invariant spectrum. After inflation, one draws $\delta({\bf k})$ from $P_{\mbox{\tiny{inf}}}(k)$, as well as photons, electron fields etc, and evolves with a program such as CMBFAST or CAMB to sometime after recombination. This should be adequately captured by the linear evolution, but should be a fully relativistic calculation. 

Lets call the initial scale factor after inflation $a_i$ and the late time to which we evolve under linear evolution by $a_{late}$. The final density fluctuation will be related to the initial density fluctuation by the transfer function $T(k)$ and the growth function $D(a)$.
One defines the transfer function as the ratio of the gravitational potentials on a given scale to that on the large scales, i.e.,
\beq
T(k)={\phi(k,a_{late})\over \phi_{ls}(k,a_{late})}
\eeq
In particular for modes that enter during the matter dominated era, we have $T(k\lesssim k_{eq})\approx1$. The transfer function decreases for modes that entered in the radiation dominated era. In particular
\beq
T(k)\approx{12 k_{eq}^2\over k^2}\ln\!\left({k\over 8 k_{eq}}\right),\,\,\,\,\,\,k\gg k_{eq}
\eeq
with wavenumber at equality \cite{Dodelson}
\beq
k_{eq}\approx 0.073\,\mbox{Mpc}^{-1}\,\Omega_{m,0}h^2
\eeq
Plus, there are corrections from baryon-acoustic-oscillations etc on the spectrum which CMBFAST or CAMB should provide.

At the linear level,  we must then simply multiply by the growth function $D(a)$ to reach today's spectrum. We express the evolution in the Newtonian potential as follows
\beq
\phi_L(k,a)={9\over 10}\phi_{\mbox{\tiny{inf}}}(k) T(k) {D(a)\over a}
\eeq
where the $9/10$ prefactor is from the Sachs-Wolfe effect, $\phi_{\mbox{\tiny{inf}}}$ is the primordial fluctuation, $T(k)$ is the transfer function,
and $D(a)$ is the growth factor normalized to the scale factor $a$ for convenience. Later we will generalize the growth factor from $D(a)\to D(k,a)$ to account for the $k$-dependence in the resummed linear theory, but lets suppress that for now.

\begin{figure}[t]
\center{\includegraphics[width=11cm]{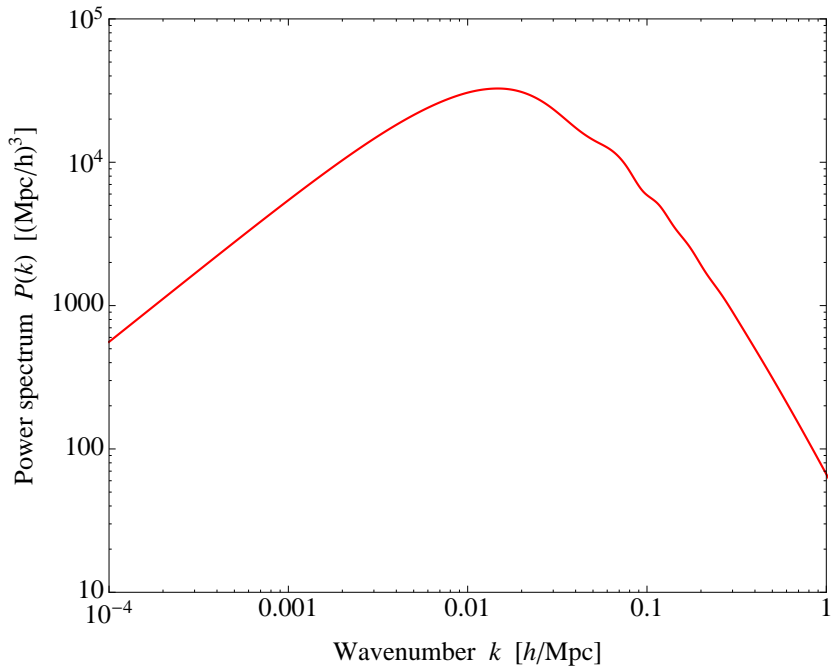}}
\caption{Linear power spectrum of density fluctuations $P_L(k)$ computed from CAMB, with $n_s=0.96$, $z=0$, $\Omega_m=0.226$, $\Omega_k=0$. The plot for $k<k_{eq}$ shows the approximate scale invariance of the spectrum.}
\label{CAMBa}\end{figure}

\begin{figure}[t]
\center{\includegraphics[width=11.3cm]{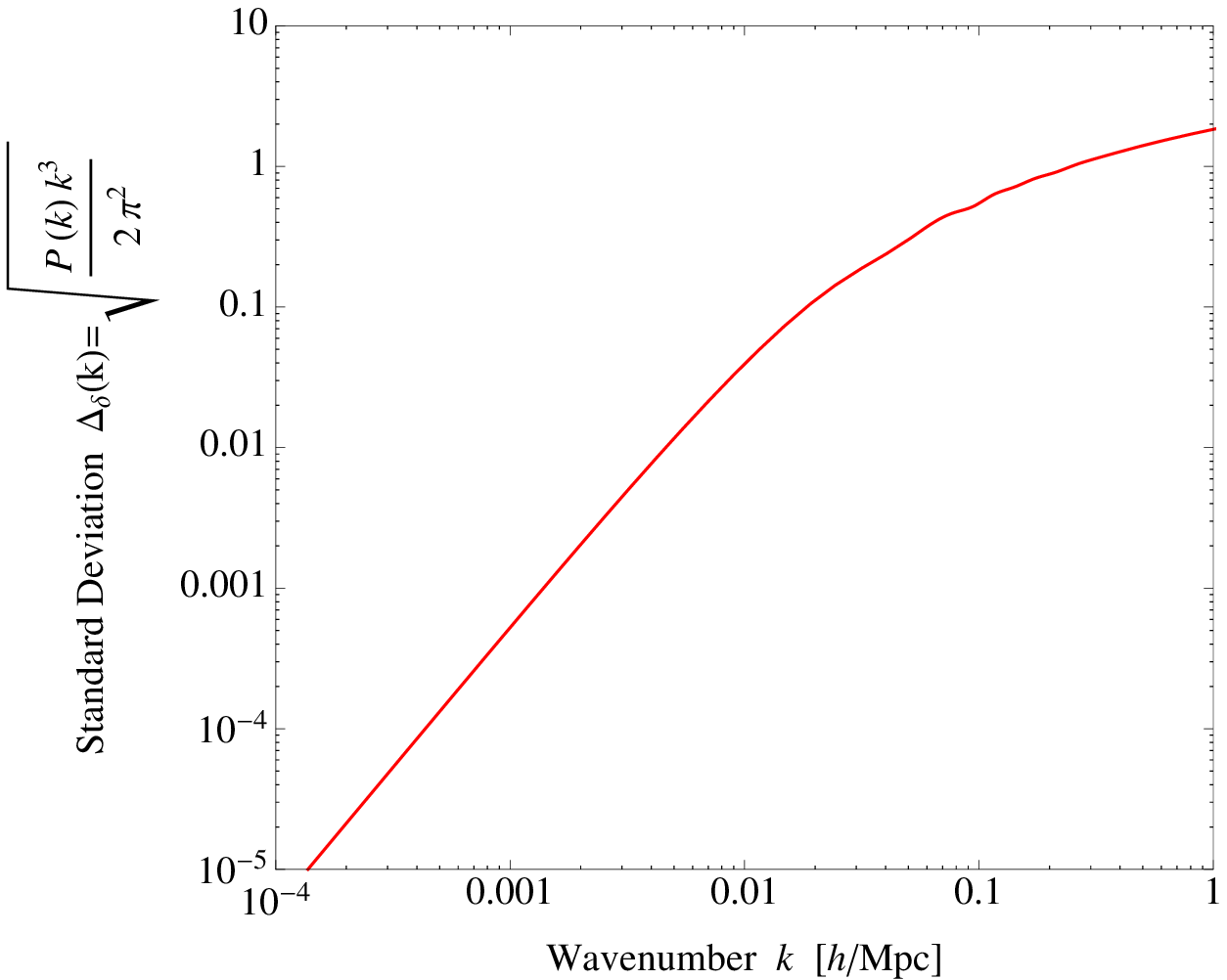}}
\caption{Linear standard deviation of density fluctuations  $\Delta_\delta(k)$ computed from CAMB, with $n_s=0.96$, $z=0$, $\Omega_m=0.226$, $\Omega_k=0$. The plot indicates that the evolution is perturbative for small $k$ and non-perturbative for high $k$.}
\label{CAMBb}\end{figure}

For sub-horizon modes, the Poisson equation gives the following relationship between density and Newtonian potential
\beq
\delta_L={2k^2\phi_L \, a\over 3 \Omega_{m,0}H_0^2}
\eeq
where $\Omega_{m,0}$ and $H_0$ are today's values.
So the density fluctuation is given in terms of the primordial fluctuations, transfer function, and growth factor by
\beq
\delta_L(k,a)={3\over 5}{k^2\over\Omega_{m,0}H_0^2}\phi_{\mbox{\tiny{inf}}}(k)T(k)D(a)
\eeq
The primordial power spectrum generated during inflation is 
\beq
P_\phi(k)={8\pi G\over 9 k^3}{H_{\mbox{\tiny{inf}}}^2\over\epsilon}
\eeq
For a power law, we write the power spectrum as
\beq
P_\phi(k)={50\pi^2\over 9 k^3}\left(k\over H_0\right)^{n_s-1}\delta_H^2\left({\Omega_{m,0}\over D(a=1)}\right)^2
\eeq
where we followed \cite{Dodelson} in the definition of the amplitude, denoted $\delta_H$.
Combining the above, we have the linear power spectrum in the density as
\beq
P_L(k,a)=2\pi^2\delta_H^2{k^{n_s}\over H_0^{n_s+3}}T^2(k)\left({D(a)\over D(a=1)}\right)^2
\eeq
From running CAMB, a plot of this linear power spectrum is given in Fig.~\ref{CAMBa}. 
The dimensionless variance is defined as
\beq
\Delta_\delta^2(k,a)\equiv{k^3 P_L(k,a)\over 2\pi^2}
\label{Variance}\eeq
Note that on Hubbles scales today; $\Delta_\delta^2(H_0,a=1)=\delta_H^2$, which explains the unconventional normalization chosen in $P_\phi(k)$. The measured value of the amplitude of density fluctuations on the scale of the horizon is $\delta_H\approx 1.9\times 10^{-5}$ \cite{Komatsu:2010fb}.

The standard deviation $\Delta_\delta(k,a)$ is a measure of the fluctuations in $\delta$ on a scale $k$, which we plot in Fig.~\ref{CAMBb}. Since these fluctuations become larger than 1 at high $k$, the theory is non-linear in this regime, which sets a non-linear scale of $\knl\sim 0.5$ [h/Mpc], or so. This defines the perturbative regime in which the effective fluid description is applicable for $k<\knl$ and the non-perturbative regime in which the effective fluid description breaks down for $k>\knl$. The scale $\knl$ acts a type of ``coupling" in the effective field theory and organizes the expansion into powers of $k/\knl$.

\begin{figure}[t]
\center{\includegraphics[width=11cm]{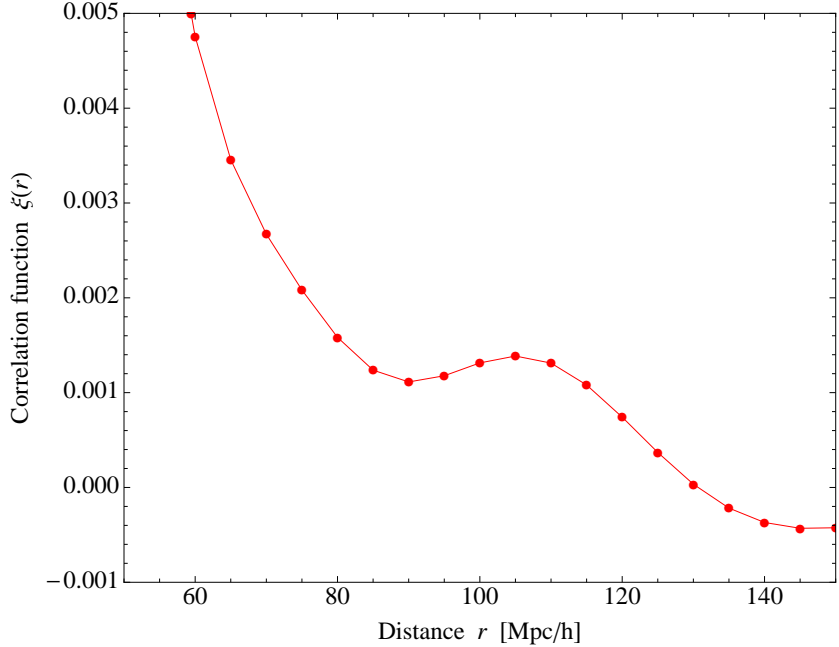}}
\caption{Linear density correlation function $\xi(r)$ computed from CAMB 
in a $\Lambda$CDM universe, with $n_s=0.96$, $z=0$, $\Omega_m=0.226$, $\Omega_k=0$.
This clearly shows  the baryon-acoustic-oscillation peak at $r\sim 120$\,[Mpc/h].}
\label{CAMB2}\end{figure}

By reverting back to position space we can define the correlation function $\xi$.
Using statistical homogeneity and isotropy it is related to the variance by
\bea
\xi(r,a)\equiv\langle\delta({\bf x},a)\,\delta({\bf x}+{\bf r},a)\rangle
=\int\! d\ln k\,{\sin(kr)\over kr}\Delta_\delta^2(k,a)
\eea
and is in Fig.~\ref{CAMB2}. The correlation function evidently includes the baryon-acoustic-oscillation peak. Its precise shape is subject to non-linearities that we would like to compute accurately; although this is analyzed most cleanly in $k$-space, which will be our focus.

\subsection{Evolution Equations}

In order to study non-linear corrections, we begin by recalling here our equation of motion for the velocity field for the stress-tensor ansatz we made earlier
\bea
\dot v_l^i+H v_l^i+v_l^j\partial_jv_l^i+{1\over a}\partial_i\phi_l=
-{1\over a}c_s^2\,\partial_i\delta_l
+{3c_{sv}^2\over4Ha^2}\partial^2v_l^i+{4c_{bv}^2+c_{sv}^2\over 4Ha^2}\partial_i\partial_jv_l^j
-\Delta J^i\,\,\label{eulertot}
\eea
where $\Delta J^i\equiv \rho_b^{-1}\partial_j\Delta\tau^{ij}/a$. Which is complimented by the Poisson and continuity equations.
Our ansatz for $[\tau^{ij}]_\Lambda$ suggests that the right hand side of (\ref{eulertot}) should be multiplied by an overall prefactor $1/(1+\delta_l)$. But since we have only included linear terms in our ansatz for $[\tau^{ij}]_\Lambda$ it would not be self-consistent to include this prefactor. Furthermore, the leading order from such terms would enter parametrically at 3rd order in an expansion in powers of the linear density field $\delta^{(1)}$ as $c_s^2\nabla^2\delta^{(1)}\,\delta^{(1)}$. This would correct the two-point correlation function as $\langle c_s^2\nabla^2\delta^{(1)}\,\delta^{(1)}\,\delta^{(1)}\rangle$, which obviously vanishes when the primordial spectrum is Gaussian (or just even in $\delta^{(1)}$). Hence we can drop such corrections.

\subsubsection{Curl of Velocity}
\begin{figure}[t]
\center{\includegraphics[width=11cm]{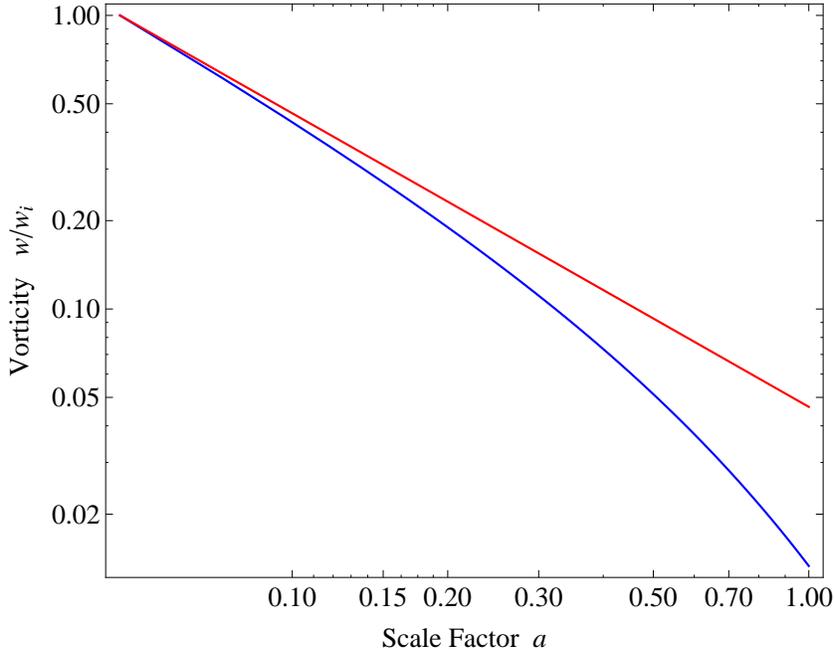}}
\caption{Curl of velocity ${\bf w}$ (normalized to some initial value) in a $k$-mode as a function of time in a matter dominated era in the linear approximation.
Blue is for $c_{sv}^2>0$ with re-summation and red is for $c_{sv}^2=0$. This shows the decay in vorticity over time at this order of analysis.}
\label{FirstOrderVelCurl}\end{figure}

Before examining the density fluctuations in detail, let us briefly mention the vorticity.
The curl, or vorticity, of the velocity field
\beq
{\bf w}_l\equiv\nabla\times{\bf v}_l\,\,\,\,\,\mbox{or in $k$-space}\,\,\,\,{\bf w}_l\equiv i\,{\bf k}\times{\bf v}_l
\eeq
is determined by taking the curl of the Euler equation. We use the vector identity
\beq
\nabla\times({\bf v}_l\cdot\nabla){\bf v}_l)=-\nabla\times({\bf v}_l\times(\nabla\times{\bf v}_l))
\eeq
to obtain the non-linear vorticity equation 
\bea
\left({d\over dt}+H-{3c_{sv}^2\nabla^2\over 4Ha^2}\right){\bf w}_l=\nabla\times\left({1\over a}{\bf v}_l\times{\bf w}_l-\Delta{\bf J}\right)\,\,\,\,\,
\eea
Let's first discuss this at the linear level ${\bf w}^{(1)}$ where we ignore the right hand side.
Even in the absence of viscosity, the vorticity ${\bf w}^{(1)}$ is being driven to zero in an expanding universe as ${\bf w}_l\propto 1/a$, which is a well known result.
In the presence of viscosity, this happens all the more rapidly (assuming $c_{sv}^2>0$). The curl of velocity in a matter dominated universe is plotted in Figure \ref{FirstOrderVelCurl}. This means that at the linear level, studied at late times, we cannot see the effect of the shear-viscosity as the transient vorticity will have decayed away. 
The non-linear term on the right hand side also vanishes when ${\bf w}_l=0$ and therefore vorticity is not generated, unless it is present initially (although the curl of the stochastic fluctuations $\nabla\times\Delta{\bf J}$ could alter this). 

However, other non-linear terms that we have neglected, such as the overall prefactor $1/(1+\delta_l)$, will generate vorticity.
These are required for the shear viscosity to have an explicit effect, if we cannot track the initial transient viscosity.  On the other hand, the combination of viscosities $c_v^2=c_{sv}^2+c_{bv}^2$ will appear in the divergence of the Euler equation (see eq.~(\ref{Eulertau})) and can play an important role at late times. For the present discussion, let us assume the vorticity is negligible. This is consistent with simulations and observations which find the vorticity to be small, albeit non-zero.

\subsubsection{Divergence of Velocity}

The divergence of the velocity 
\beq
\theta_l\equiv\nabla\cdot{\bf v}_l\,\,\,\,\,\mbox{or in $k$-space}\,\,\,\,\theta_l\equiv i\,{\bf k}\cdot{\bf v}_l
\eeq
is coupled to the density fluctuation $\delta_l$. Lets write down the coupled equations using conformal time 
and the associated Hubble parameter
\beq
\tau=\int\!{dt\over a(t)}\,\,\,\,\,\,\,\,\,\,\mathcal{H}={1\over a}{da\over d\tau}
\eeq
In the absence of vorticity, and ignoring stochastic fluctuations for now (see Section \ref{StochasticFluctuations} for its inclusion) the evolution equations for the pair $\delta_l,\,\theta_l$ are found to be
\bea
{d\delta_l\over d\tau}+\theta_l\amp=\amp -\int\!{d^3k'\over(2\pi)^3}\alpha({\bf k},{\bf k}')\delta_l({\bf k}-{\bf k}')\theta_l({\bf k}')\,\,\,\,\,\,\label{Conttau}\\
{d\theta_l\over d\tau}+\mathcal{H}\theta_l+{3\over2}\mathcal{H}^2\Omega_m\delta_l
\amp=\amp-\int\!{d^3k'\over(2\pi)^3}\beta({\bf k},{\bf k}')\theta_l({\bf k}-{\bf k}')\theta_l({\bf k}')
+c_s^2k^2\delta_l-{c_v^2k^2\over\mathcal{H}}\theta_l\label{Eulertau}
\eea
which comes from taking a divergence of the Euler equation.
Here
\bea
\alpha({\bf k},{\bf k}')\amp\equiv\amp{{\bf k}\cdot{\bf k}'\over (k')^2}\\
\beta({\bf k},{\bf k}')\amp\equiv\amp{k^2\,{\bf k}'\cdot({\bf k}-{\bf k}')\over 2|{\bf k}'|^2|{\bf k}-{\bf k}'|^2}
\eea
appear as the kernels of the above convolution integrals. The convolution integrals on the right hand side of eqs.~(\ref{Conttau}--\ref{Eulertau}) arise from long-long mode coupling, while the $c_s$ and $c_v$ terms arise from long-short mode coupling.

\subsection{Recursion Relations}

Since we will be evolving the universe fully linearly during the radiation dominated era, until after decoupling, the subsequent evolution will be during a matter dominated era. Although the late time behavior with a cosmological constant will alter this quantitatively, let us ignore this for now and study the matter dominated era for analytical simplicity; the generalization to the $\Lambda$CDM case will be performed in Section \ref{LCDM}. In the matter dominated case we have
\beq
\Omega_m=1,\,\,\,\,\mathcal{H}={2\over\tau},\,\,\,\,a=\left({\tau\over\tau_0}\right)^2
\eeq
Given this, we would like to extract the time dependence in the problem, by performing a self consistent expansion in the scale factor, as Hubble is the only time scale in the problem. 

Recently in Ref.~\cite{Carrasco:2012cv} we exploited the use of Green's functions to capture the time dependence, which is a very powerful technique. This led to result in terms of integrals over factors of Green's functions, etc. Here we would like to extract the time dependence in a more explicit and intuitive way to compliment the previous powerful results.
Since the density field and velocity field is small at early times, we expand our fields in powers of the scale factor as follows
\bea
\delta_l({\bf k},\tau)\amp=\amp\sum_{n=1}^\infty a^n(\tau)\delta_n({\bf k})\label{expdelta}\\
\theta_l({\bf k},\tau)\amp=\amp-\mathcal{H}\sum_{n=1}^\infty a^n(\tau)\theta_n({\bf k})
\eea
where we have suppressed ``l" subscripts in the perturbed expansion on the RHS, although they are all long-modes.
This cleanly separates out the time dependence and the $k$-dependence.
Ordinarily, the higher order contributions to the density fields $\delta_n$ can be counted in powers of the linearized density fluctuation $\delta_1$, namely $\delta_n\sim\theta_n\sim\delta_1^n$. 
However, the presence of the fluid corrections alters this simple counting here.
Note that the leading order terms satisfy $\delta_1=\theta_1$, with growth factor $D(a)=a$.
The linear density fluctuation $\delta_1({\bf k})$ is therefore drawn from the time independent power spectrum
\bea
P_{11}(k)\amp\equiv\amp{P_L(k,a)\over a^2}W_\Lambda(k)^2\\
\amp=\amp 2\pi^2\delta_H^2{k^{n_s}\over H_0^{n_s+3}}T^2(k)W_\Lambda(k)^2\label{P11W}
\eea
where the factor $W_\Lambda(k)^2$ comes from smoothing
 and $T(k)$ is the transfer function that includes the full effects from the radiation dominated era as computed by a program such as CMBFAST or CAMB.

If we substitute the expansions for $\delta_l$ and $\theta_l$ into the continuity and Euler equations, we can equate powers of the scale factor $a$, giving a pair of recursion relations for $\delta_n$ and $\theta_n$. This will depend on the time dependence of the fluid parameters $c_s$ and $c_v$. In a matter dominated era, we will show in the Section \ref{Timedepfluid}, that the approximate time dependence of the fluid parameters is that they increase with the scale factor $a$. To capture this we introduce
\beq
c_s^2(a,\Lambda)=a\,c_{s,0}^2(\Lambda),\,\,\,\,\,\,\,\,c_v^2(a,\Lambda)=a\,c_{v,0}^2(\Lambda)\label{csvtime}
\eeq
and we define the following time independent  dimensionless  parameters
\bea
C_{s,0}(k)\equiv{c_{s,0}^2 k^2\over\mathcal{H}_0^2},\,\,\,\,\,\,\,\,\,C_{v,0}(k)\equiv{c_{v,0}^2 k^2\over\mathcal{H}_0^2}
\eea
where the $\Lambda$ dependence is implied.
For $n>1$ we find the following set of  relationships between the fields at different orders
\bea
A_n({\bf k})\amp=\amp n\,\delta_n({\bf k})-\theta_n({\bf k})\\
B_n({\bf k})\amp=\amp 3\delta_n({\bf k})-(2n+1)\theta_n({\bf k})-2C_{s,0}(k)\delta_{n-2}({\bf k})
-2C_{v,0}(k)\theta_{n-2}({\bf k})
\eea
where the left hand side are the following non-linear integrals
\bea
A_n({\bf k})\amp=\amp\int{d^3k_1\over(2\pi)^3}\int d^3k_2\, \delta_D^3({\bf k}_1+{\bf k}_2-{\bf k})\nonumber\\
\amp\amp\alpha({\bf k},{\bf k}_1)\sum_{m=1}^{n-1}\theta_m({\bf k}_1)\delta_{n-m}({\bf k}_2)\\
B_n({\bf k})\amp=\amp-\int{d^3k_1\over(2\pi)^3}\int d^3k_2\, \delta_D^3({\bf k}_1+{\bf k}_2-{\bf k})\nonumber\\
\amp\amp2\beta({\bf k},{\bf k}_1)\sum_{m=1}^{n-1}\theta_m({\bf k}_1)\theta_{n-m}({\bf k}_2)
\eea
These  relationships allow us to express the $m=n^{th}$ value of the fields in terms of the $m<n^{th}$ value of the fields. Namely, we have the following recursion relations for $n>1$
\bea
\delta_n({\bf k})\amp=\amp{1\over(2n+3)(n-1)}\Big{[}
(2n+1)A_n({\bf k})-B_n({\bf k})\nonumber\\
\amp\amp-2C_{s,0}(k)\delta_{n-2}({\bf k})-2C_{v,0}(k)\theta_{n-2}({\bf k})\Big{]}\label{recdelta}\\
\theta_n({\bf k})\amp=\amp{1\over(2n+3)(n-1)}\Big{[}
3A_n({\bf k})-nB_n({\bf k})\nonumber\\
\amp\amp-2nC_{s,0}(k)\delta_{n-2}({\bf k})-2nC_{v,0}(k)\theta_{n-2}({\bf k})\Big{]}\label{rectheta}
\eea
with starting values $\theta_1({\bf k})=\delta_1({\bf k})$. This allows us to in principle solve for all the higher order fields in terms of $\delta_1({\bf k})$.

The solution for $\delta_n$ and $\theta_n$ can be expressed in terms of kernels $F_{n,j}$ and $G_{n,j}$, rather than left in terms of the stochastic variable $\delta_1$. Lets write our fields as
\bea
\delta_n({\bf k})\amp=\amp\sum_{j=1}^n\int\!{d^3q_1\over(2\pi)^3}\ldots\int\!d^3q_j\,\delta_D^3({\bf q}_1+\cdots+{\bf q}_j-{\bf k})\nonumber\\
\amp\amp\,\,\,\,\, F_{n,j}({\bf q}_1,\ldots,{\bf q}_j)\,\delta_1({\bf q}_1)\ldots\delta_1({\bf q}_j)\\
\theta_n({\bf k})\amp=\amp\sum_{j=1}^n\int\!{d^3q_1\over(2\pi)^3}\ldots\int\!d^3q_j\,\delta_D^3({\bf q}_1+\cdots+{\bf q}_j-{\bf k})\nonumber\\
\amp\amp\,\,\,\,\, G_{n,j}({\bf q}_1,\ldots,{\bf q}_j)\,\delta_1({\bf q}_1)\ldots\delta_1({\bf q}_j)
\eea
Then by substituting into (\ref{recdelta}--\ref{rectheta}) we establish recursion relations for $F_{n,j}$ and $G_{n,j}$ with starting values $F_{1,1}=G_{1,1}=1$. At the one-loop order, we will find that only $F_{n,n}$, $G_{n,n}$ and $F_{n,1}$, $G_{n,1}$ will enter, if the primordial fluctuations are Gaussian. So we report on their values here. Firstly, $F_{n,n}$ and $G_{n,n}$ are independent of the fluid parameters and are given recursively by
\bea
F_{n,n}({\bf q}_1,\ldots,{\bf q}_n)\amp=\amp\sum_{m=1}^{n-1}{G_{m,m}({\bf q}_1,\ldots,{\bf q}_m)\over(2n+3)(n-1)}\times\nonumber\\
\amp\amp\Big{[}(2n+1)\alpha({\bf k},{\bf k}_1)F_{n-m,n-m}({\bf q}_{m+1},\ldots,{\bf q}_n)\nonumber\\
\amp\amp+2\beta({\bf k}_1,{\bf k}_2)G_{n-m,n-m}({\bf q}_{m+1},\ldots,{\bf q}_n)\Big{]}\\
G_{n,n}({\bf q}_1,\ldots,{\bf q}_n)\amp=\amp\sum_{m=1}^{n-1}{G_{m,m}({\bf q}_1,\ldots,{\bf q}_m)\over(2n+3)(n-1)}\times\nonumber\\
\amp\amp\Big{[}3\alpha({\bf k},{\bf k}_1)F_{n-m,n-m}({\bf q}_{m+1},\ldots,{\bf q}_n)\nonumber\\
\amp\amp+2n\beta({\bf k}_1,{\bf k}_2)G_{n-m,n-m}({\bf q}_{m+1},\ldots,{\bf q}_n)\Big{]}
\eea
On the other hand, $F_{n,1}(k)$ and $G_{n,1}(k)$ are determined entirely by the fluid parameters. We find them to be the following products ($n>1$)
\bea
F_{n,1}(k)\amp=\amp\prod_{m=3,5,\ldots}^n{-2(C_s(k)+(m-2)C_v(k))\over(2m+3)(m-1)},\,\,\,\,\,\,\,\,\,\,\,\,\,\mbox{for}\,\,\,n\,\,\,\mbox{odd}\\
G_{n,1}(k)\amp=\amp n\prod_{m=3,5,\ldots}^n{-2(C_s(k)+(m-2)C_{v,0}(k))\over(2m+3)(m-1)},\,\,\,\,\,\mbox{for}\,\,\,n\,\,\,\mbox{odd}
\eea
and we find $F_{n,1}=G_{n,1}=0$ for $n$ even.
Note that $G_{n,1}=n \, F_{n,1}$.

\subsection{Power Spectrum}\label{PSpectrum}

We will go to one-loop order in the power spectrum. This will require the second and third order density corrections.
At second order, we have the following (symmetrized) kernels
\bea
F_{2,2}^{(s)}({\bf k}_1,{\bf k}_2)\amp=\amp{5\over7}+{2\over7}{({\bf k}_1\cdot{\bf k}_2)^2\over k_1^2 k_2^2}+{{\bf k}_1\cdot{\bf k}_2\over 2}\left({1\over k_1^2}+{1\over k_2^2}\right)\,\,\,\,\,\,\,\,\,\,\,\,\\
G_{2,2}^{(s)}({\bf k}_1,{\bf k}_2)\amp=\amp{3\over7}+{4\over7}{({\bf k}_1\cdot{\bf k}_2)^2\over k_1^2 k_2^2}+{{\bf k}_1\cdot{\bf k}_2\over 2}\left({1\over k_1^2}+{1\over k_2^2}\right)\,\,\,\,\,\,\,\,\,\,\,\,\\
F_{2,1}(k)\amp=\amp0\\
G_{2,1}(k)\amp=\amp0
\eea
At third order we need $F_{3,3}$ , $G_{3,3}$, $F_{3,1}$, and $G_{3,1}$ which we find to be the following (unsymmetrized) kernels
\bea
F_{3,3}({\bf q}_1,{\bf q}_2,{\bf q}_3) \amp=\amp{1\over18}\Big{[}7\alpha({\bf k},{\bf q}_1)F_{2,2}({\bf q}_2,{\bf q}_3)+2\beta({\bf q}_1,{\bf q}_2+{\bf q}_3)G_{2,2}({\bf q}_2,{\bf q}_3)\nonumber\\
\amp\amp+ (7\alpha({\bf k},{\bf q}_1+{\bf q}_2)+2\beta({\bf q}_1+{\bf q}_2,{\bf q}_3)G_{2,2}({\bf q}_1,{\bf q}_2)\Big{]}\,\,\,\,\,\,\,\,\,\,\,\,\,\\
 G_{3,3}({\bf q}_1,{\bf q}_2,{\bf q}_3)\amp=\amp{1\over18}\Big{[}3\alpha({\bf k},{\bf q}_1)F_{2,2}({\bf q}_2,{\bf q}_3)+6\beta({\bf q}_1,{\bf q}_2+{\bf q}_3)G_{2,2}({\bf q}_2,{\bf q}_3)\nonumber\\
\amp\amp+ (3\alpha({\bf k},{\bf q}_1+{\bf q}_2)+6\beta({\bf q}_1+{\bf q}_2,{\bf q}_3)G_{2,2}({\bf q}_1,{\bf q}_2)\Big{]}\,\,\,\,\,\,\,\,\,\,\,\,\,\\
F_{3,1}(k) \amp=\amp
 -{1\over 9}(C_{s,0}(k)+C_{v,0}(k))\\
 G_{3,1}(k) \amp=\amp
- {1\over 3}(C_{s,0}(k)+C_{v,0}(k))
\eea

The two-point function for $\delta_l$ defines the smoothed power as follows
\beq
\langle\delta_l({\bf k},\tau)\delta_l({\bf k}',\tau)\rangle=(2\pi)^3\delta_D^3({\bf k}+{\bf k}')P(k,\tau)
\eeq
We now substitute in the expansion (\ref{expdelta}) for $\delta_l$ in powers of the scale factor to give the two-point function the form
\bea
\langle\delta_l({\bf k},\tau)\delta_l({\bf k}',\tau)\rangle \amp=\amp
 a^2(\tau)\langle\delta_1({\bf k})\delta_1({\bf k})\rangle+2a^3(\tau)\langle\delta_1({\bf k})\delta_2({\bf k}')\rangle\nonumber\\
\amp+\amp a^4(\tau)\left[2\langle\delta_1({\bf k})\delta_3({\bf k}')\rangle+\langle\delta_2({\bf k})\delta_2({\bf k}')\rangle\right]+\ldots\,\,\,\,\,\,\,\,\,\,\,\,\,\,\,\,\label{2ptexp}
\eea
In order to organize this into an expansion in power spectra, let us define the power spectrum that contributes at $n^{th}$ order as
\beq
\langle\delta_m({\bf k})\delta_{n-m}({\bf k}')\rangle=(2\pi)^3\delta_D({\bf k}+{\bf k}')P_{m\,n-m}(k,\tau)
\eeq
By inserting this into (\ref{2ptexp}) we obtain the following expansion for the power spectrum
\bea
P(k,\tau)\amp=\amp a^2(\tau)P_{11}(k)+2a^3(\tau)P_{12}(k)\nonumber\\
\amp+\amp a^4(\tau)\left[2P_{13}(k)+P_{22}(k)\right]+\ldots
\label{Pexp}\eea

\begin{figure}[h]
\center{\includegraphics[width=9.6cm]{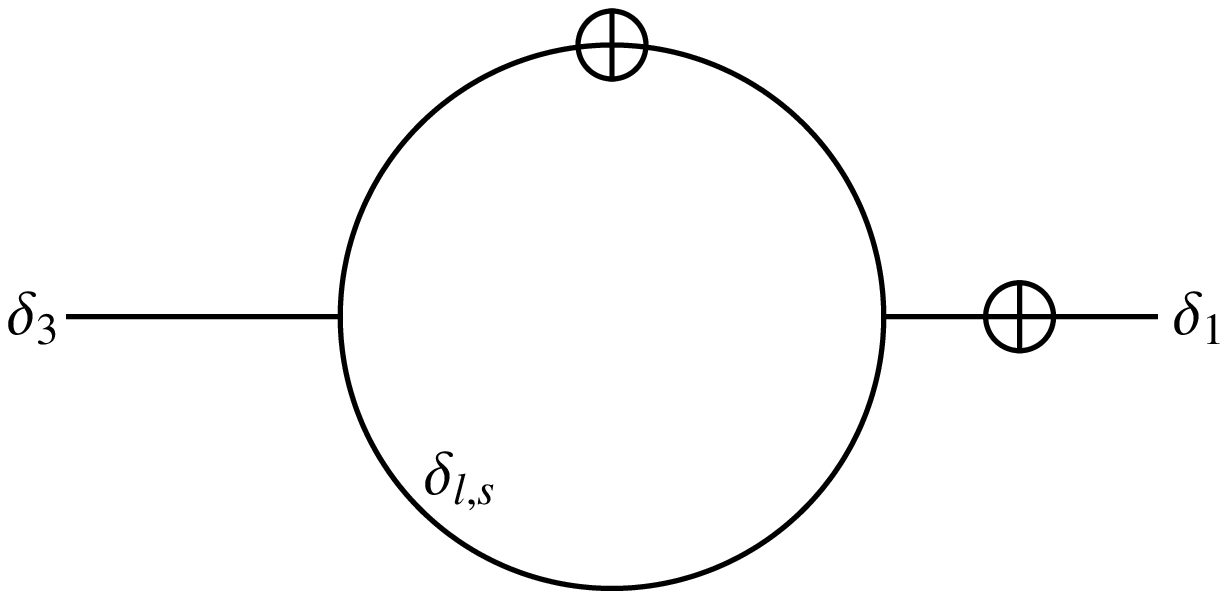}\,
\includegraphics[width=9.6cm]{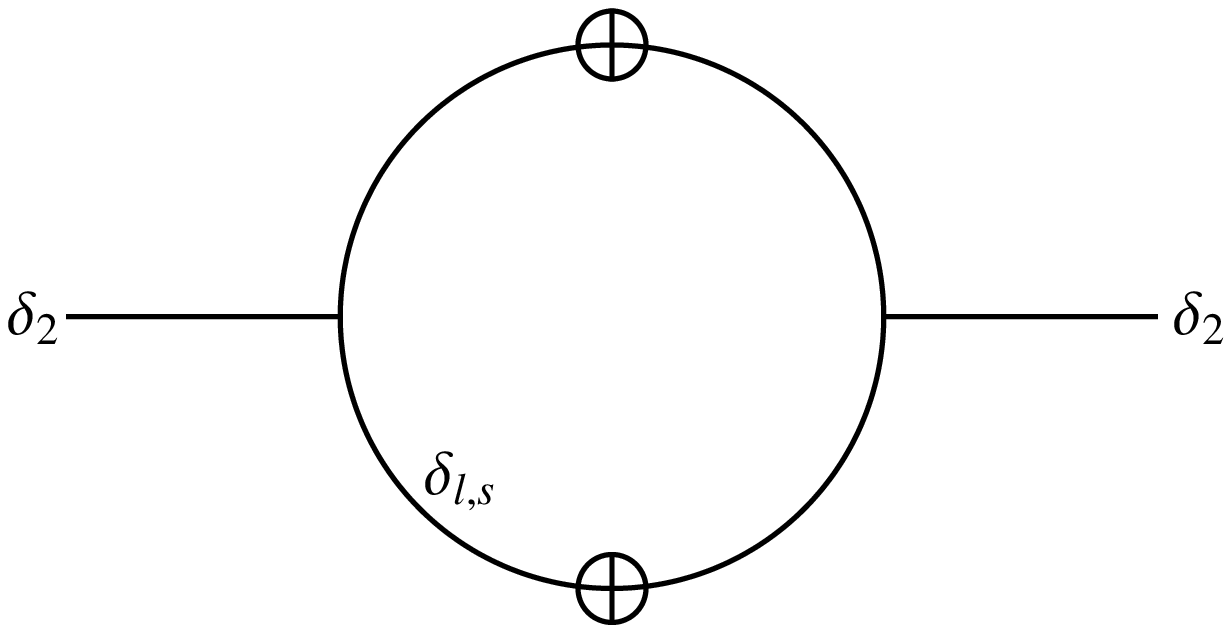}}
\caption{One loop Feynman diagrams for $P_{13}$ (top) and $P_{22}$ (bottom) after having extracted the time dependence. The crossed circles represent an insertion of the linear power spectrum and the loops represent convolution integrals cutoff at $q\sim\Lambda$.}
\label{FeynmanDiagrams}\end{figure}

We assume that the primordial power spectrum is Gaussian, allowing us to simplify this expansion.
This implies that
\beq
P_{12}=0, 
\eeq
due to $\delta_2$ being symmetric under $\delta_1\to-\delta_1$.
At the next order, we find various contributions including the four-point function of $\delta_1$, which can be simplified using Wick's theorem. The final result for $P_{13}$ and $P_{22}$ will in general have two contributions: the contributions from IR modes and the contribution from UV modes, which we write in the following obvious notation
\bea
P_{13}(k)=P_{13,IR}(k,\Lambda)+P_{13,UV}(k,\Lambda)\\
P_{22}(k)=P_{22,IR}(k,\Lambda)+P_{22,UV}(k,\Lambda)
\eea
where we have reinstated the $\Lambda$ dependence on the RHS as it separates the IR and the UV modes.
By the IR contribution we mean the usual one-loop contribution, but cutoff at wavenumber $\Lambda$; this is given by the following loop integrals
\bea
P_{13,IR}(k,\Lambda)\amp=\amp 3\,P_{11}(k)\int^\Lambda\!\!{d^3q\over(2\pi)^3}F_{3,3}^{(s)}({\bf q},-{\bf q},{\bf k})P_{11}(q)\label{P13IR}\\
P_{22,IR}(k,\Lambda)\amp=\amp 2\int^\Lambda\!\!{d^3q\over(2\pi)^3}\left[F_{2,2}^{(s)}({\bf q},{\bf k}-{\bf q})\right]^2\!P_{11}(q)P_{11}(|{\bf k}-{\bf q}|)\label{P22IR}
\eea
These contributions have a Feynman diagram representation that we present in Fig.~\ref{FeynmanDiagrams}.
We have put a $\Lambda$ superscript on the integrals as a reminder that they are cutoff by the smoothing function $W_\Lambda(k)$ that appears in $P_{11}$.
By the UV contribution we mean the new fluid contribution that we are for the first time including in this work. This is given by the following
\bea
P_{13,UV}(k,\Lambda)\amp=\amp F_{3,1}(k,\Lambda)P_{11}(k)\nonumber\\
\amp=\amp -{(c_{s,0}^2(\Lambda)+c_{v,0}^2(\Lambda))k^2\over9\mathcal{H}_0^2}P_{11}(k)\label{P13UV}\\
P_{22,UV}(k,\Lambda)\amp=\amp\Delta P_{22}(k,\Lambda)\label{P22UV}
\eea
Here $P_{13}$ is set by the ($\Lambda$ dependent) sound speed and viscosity, and
$\Delta P_{22}$ is set by the stochastic fluctuations that we elaborate on in Section \ref{StochasticFluctuations}; the latter we find to be smaller than the former at low $k$ as there is a suppression in the UV part of the integral by the transfer function. The IR contributions are associated with the long modes $\delta_l$ running in the loop, while the UV contributions are associated with the short modes $\delta_s$ running in the loop.

\subsection{Cutoff Dependence of Fluid Parameters}\label{Timedepfluid}

For the cutoff in the perturbative regime ($\Lambda\lesssim \knl$), the $\Lambda$ dependence of the fluid parameters is adequately described by the linear theory. This allows us to estimate the value of the fluid parameters $c_s^2$ and $c_v^2$ and their time dependence as a function of the linear power spectrum. The sound speed is roughly given by the velocity dispersion, so in linear theory we estimate the sound speed by an integral over the velocity dispersion of the short modes. The linear theory is not applicable at very high $k$, so we shall include a constant correction as follows
\beq
c_s^2(a,\Lambda)=\alpha \int_\Lambda d\ln q\,\,\Delta_v^2(q)+c_s^2(a,\infty)
\label{csint}\eeq
where $\Delta_v^2$ is the velocity dispersion, 
$\alpha$ is an $\mathcal{O}(1)$ constant of proportionality (which we will fix later),
and $\int_\Lambda d\ln q\equiv \int d\ln q\, (1-W_\Lambda(k))^2$ since the sound speed  arises from integrating out the {\em short} modes.
In the $\Lambda\to\infty$ limit, which we shall eventually take once we cancel the $\Lambda$ dependence, we find that $c_s^2(\Lambda)$ is non-zero (due to the UV dependence), which we account for with the $c_s^2(\infty)$ constant correction.

Now we would like to relate the velocity to the power spectrum in terms of $\phi$ or $\delta$. In the linear theory in a matter dominated universe universe, we have the growing mode solution
\beq
{\bf v}_L={i\mathcal{H}{\bf k}\over k^2}\delta_L
\eeq
which is the first order description of modes in the perturbative regime. This includes the short modes ${\bf v}_s$ and $\delta_s$ in the regime $\Lambda<k<\knl$, which is a non-empty set if we choose small $\Lambda$.

\begin{figure}[t]
\center{\includegraphics[width=11cm]{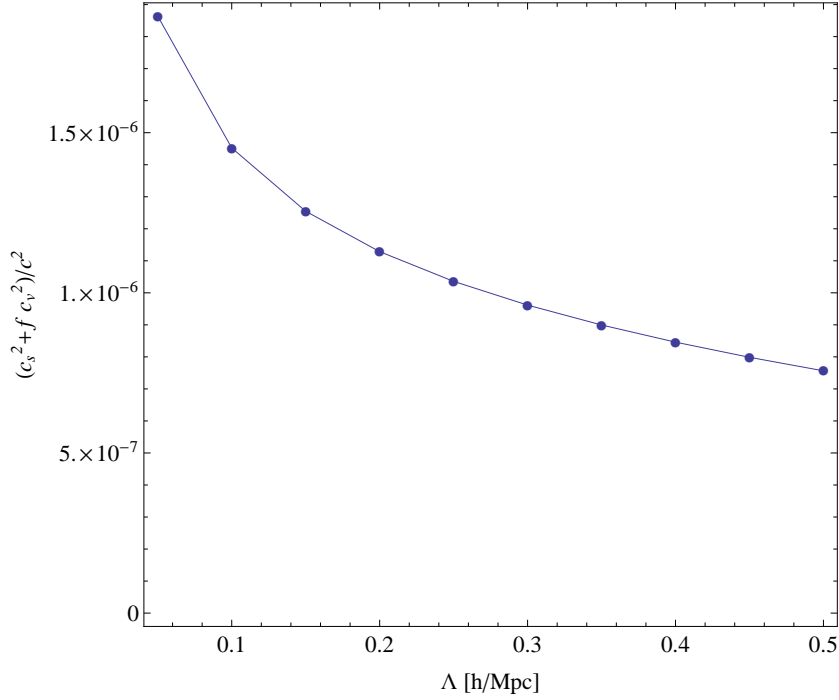}}
\caption{The (bare) fluid parameter $(c_s^2(\Lambda)+f c_v^2(\Lambda))/c^2$ at $z=0$. The $\Lambda$ dependence is chosen to cancel against the $\Lambda$ dependence of the loop integral in $P_{13}$. Note that the fluid parameter is non-zero as $\Lambda\to\infty$, which we call $c_s^2(\infty)+fc_v^2(\infty)$. The  growth parameter $f$ is different from 1 in a $\Lambda$CDM universe which is described in Section \ref{LCDM}.}
\label{csLambda}\end{figure}

This gives the following linear relationship between the variance in the velocity and the density fluctuations
\beq
\Delta_v^2(k)={\mathcal{H}^2\over k^2}\Delta_\delta^2(k)
\eeq
where $\Delta_\delta^2(k)$ is related to the density power spectrum $P_L(k)$ as given in eq.~(\ref{Variance}).
For a scale invariant primordial power spectrum we have
\bea
\Delta_\delta^2(k)\sim \Bigg{\{}
\begin{array}{c}
10^{-10}{k^4c^4\over\mathcal{H}^4},\,\,\,\,\,\,\,\,\,\,\,\,\,\,\,\,\,\,\,\,\,\,\,\,\,\,\,\,\,k\ll k_{eq}\\
10^{-8}{k_{eq}^4c^4\over\mathcal{H}^4}\ln({k\over 8 k_{eq}})^2,\,\,\,\,\,\,\,k\gg k_{eq}
\end{array}
\eea
(see Fig.~\ref{CAMBb})
where we have taken into account the transfer function which separates the modes that enter before/after matter domination.
Inserting this into eq.~(\ref{csint}) leads to the following rough estimate for the sound speed
\bea
c_s^2(a,\Lambda)\sim \Bigg{\{}
\begin{array}{c}
10^{-10}{k_{eq}^2c^4\over\mathcal{H}^2}+c_s^2(\infty),\,\,\,\,\,\,\Lambda\ll k_{eq}\\
10^{-8}{k_{eq}^4c^4\over\mathcal{H}^2\Lambda^2}+c_s^2(\infty),\,\,\,\,\,\,\,\Lambda\gg k_{eq}
\end{array}
\eea
where we have ignored logarithmic corrections, etc, so these estimates are only rough.
In order to probe baryon-acoustic-oscillations, we shall need the effective description in the regime $k\gtrsim k_{eq}$ and so we need to take $\Lambda\gg k_{eq}$ which is the latter result. 
The full result from carrying out the integral over wavenumber is given in Fig.~\ref{csLambda}, where we form a linear combination of pressure and viscosity. In the next subsection we will fix the coefficient $\alpha$ and the large $\Lambda$ asymptotic value ($c_s^2(\infty)$) that were used to produce this plot.

Notice that $c_s^2$ is time dependent, due to the $1/\mathcal{H}^2$ factor, and that it explicitly depends on the cutoff scale, due to the $1/\Lambda^2$ factor. 
A similar scaling goes through for the viscosity $c_v^2$. This leads to the $1/\mathcal{H}^2\propto a$ scaling that we stated earlier in eq.~(\ref{csvtime}).

\subsubsection{Cutoff Independence of Physical Results}\label{CutoffIndependence}

The cutoff  $\Lambda$ explicitly alters the density field $\delta_l$, velocity field $\theta_l$, etc, and hence it affects the size of the loops. It also affects the size of the (bare) fluid parameters $c_s, c_v$ etc, as summarized in Fig.~\ref{csLambda}. However, all the cutoff dependence must drop out of any physical results. In order to ensure this occurs, here we examine the form of the loop integrals (\ref{P13IR},\ref{P22IR}).

By using isotropy, the full angular integral in $P_{13,IR}$ can be performed, and the azimuthal integral can be performed in $P_{22,IR}$, this leads to
\bea
P_{13,IR}(k,\Lambda)\amp=\amp{1\over504}{k^3\over 4\pi^2}P_{11}(k)\int_0^{\Lambda/k} dr\,P_{11}(k\,r)\nonumber\\
\amp\amp\,\,\,\,\,\,\,\,\,\,\,\,\,\,\,\,\,\,\,\left({12\over r^2}-158+100r^2-42r^4+{3\over r^3}(r^2-1)^3(7r^2+2)\ln\left|{1+r\over1-r}\right|\right)\\
P_{22,IR}(k,\Lambda)\amp=\amp{1\over 98}{k^3\over 4\pi^2}\int_0^{\Lambda/k} dr\int_{-1}^1 dx\,P_{11}(k\,r)P_{11}(k\sqrt{1+r^2-2rx})    {(3r+7x-10rx^2)^2\over(1+r^2-2rx)^2}\,\,\,\,\,\,\,\,
\eea
In order to extract the $\Lambda$ dependence of $P_{13}(k,\Lambda)$ we take the large $r$ limit inside the integrand (which corresponds to the regime $k\ll\Lambda$). In this limit the integrand approaches the value $-488/5$, leading to 
\bea
P_{13,IR}(k,\Lambda)\amp=\amp P_{13,IR}(k,\Lambda_1)-{488\over5}{1\over504}{k^2\over 4\pi^2}P_{11}(k)\int^\Lambda_{\Lambda_1} dq\,P_{11}(q)
\eea
where $\Lambda_1$ is an arbitrary scale in the regime $k\ll\Lambda_1<\Lambda$. On the other hand, the UV contribution is given in eq.~(\ref{P13UV}) as
\beq
P_{13,UV}(k,\Lambda) = -{(c_{s,0}^2(\Lambda)+c_{v,0}^2(\Lambda))k^2\over9\mathcal{H}_0^2}P_{11}(k)
\eeq
Notice that the $\Lambda$ dependent parts of the IR and UV pieces have precisely the same time and $k$-dependence $\sim a^4 k^2P_{11}(k)$. Hence in order for the result to be explicitly cutoff independent, we must have
\beq
c_{s,0}^2(\Lambda)+c_{v,0}^2(\Lambda)
=\left({488\over5}{1\over504}{9\mathcal{H}_0^2\over 4\pi^2} \int_\Lambda dq\,P_{11}(q)\right)
+c_{s,0}^2(\infty)+c_{v,0}^2(\infty)
\label{csLamDep}\eeq
This fixes the coefficient $\alpha$ that we mentioned earlier at $\alpha={488\over5}{1\over504}{9\over2}$ (when combined with the viscosity term). The constant contributions are determined by explicit matching to numerical simulations.

To make this more precise, let's separate the bare fluid parameters into a renormalized part $c_{ren,0}^2(r_{ten})$ and a counter-term $c_{ctr,0}^2(k_{ren},\Lambda)$ at some renormalization scale $k_{ren}$
\beq
c_{ren,0}^2(k_{ren})+ c_{ctr,0}^2(k_{ren},\Lambda)\equiv c_{s,0}^2(\Lambda)+c_{v,0}^2(\Lambda)
\eeq
We define the counterterm to cancel the loop correction at the renormalization scale $k_{ren}$, i.e., the counterterm is defined through
\beq
P_{13,IR}(k_{ren},\Lambda)-{k_{ren}^2c_{ctr,0}^2(k_{ren},\Lambda)\over9\mathcal{H}_0^2}P_{11}(k_{ren})=0
\eeq
while the renormalized piece is defined such that the total power spectrum agrees with the full non-linear $P(k)$ result at this renormalization scale, i.e.,
\beq
P(k_{ren},\tau)=a^2(\tau)P_{11}(k_{ren})-2a^4(\tau){c_{ren}^2(k_{ren})k_{ren}^2\over 9\mathcal{H}_0^2}+a^4(\tau)P_{22}(k_{ren})
\eeq

Alternatively, by measuring the bare couplings directly from a measurement of the stress-tensor $[\tau^{ij}]_\Lambda$ in simulations, we fix $c_{s,0}^2(\Lambda)+fc_{v,0}^2(\Lambda)$ at some chosen $\Lambda$. 
The  results of this numerical work we describe in detail in Section \ref{PowerNumResults}.

\subsection{Generalization to $\Lambda$CDM}\label{LCDM}
In the previous sections we have focussed on a matter dominated era, in which case the only time scale is set by Hubble. When we include dark energy this changes the background dynamics and the evolution. Assuming the dark energy is a cosmological constant, and operating at the linear level, the velocity field is related to the density fluctuation by
\beq
{\bf v}_L={i\mathcal{H}f{\bf k}\over k^2}\delta_L
\eeq
Here $f$ is related to the  growth function 
\beq
D(a)={5\over2}H_0^2H(a)\int_0^a{da'\over(H(a')a')^3}
\eeq
by
\beq
f\equiv{d\ln D\over d\ln a}
\eeq
This can be evaluated in terms of hypergeometric functions, which we do not reproduce here.
So the corresponding relationship between the variances is
\beq
\Delta_v^2(k)={\mathcal{H}^2f^2\over k^2}\Delta_\delta^2(k)
\eeq
This leads to the fluid parameters carrying the following time dependence
\beq
c_s^2(a,\Lambda)={f^2\mathcal{H}^2D^2\over f_0^2\mathcal{H}_0^2D_0^2}c_{s,0}^2(\Lambda),\,\,\,\,\,\,
c_v^2(a,\Lambda)={f\mathcal{H}^2D^2\over f_0\mathcal{H}_0^2D_0^2}c_{v,0}^2(\Lambda)
\eeq
(where the 0 subscripts indicate the $z=0$ value, as before).

Given the numerical solution for $D$ (see the blue curve in Fig.~(\ref{FirstOrderNew})) and the time dependence of the fluid parameters, one can in principle  construct the full time dependent non-linear solution perturbatively using Green's functions; however this is quite non-trivial as the Green's function is not known. 
To make proceed, we can make use of an approximation that is known to work reasonably well; we assume that the time dependence of the $n^{th}$ order term is simply $D(k,\tau)^n$  \cite{Takahashi:2008yk}. So to generalize the expansion (\ref{expdelta}) for the matter dominated universe to the $\Lambda$CDM universe, we write
\bea
\delta_l({\bf k},\tau)\amp=\amp\sum_{n=1}^\infty D(\tau)^n\,\delta_n({\bf k})\\
\theta_l({\bf k},\tau)\amp=\amp-\mathcal{H}f\sum_{n=1}^\infty D(\tau)^n\theta_n({\bf k})
\label{resum}\eea
The approximation is finalized by taking each $\delta_n({\bf k})$ to be the value in the $D\to a$ theory, i.e., the 
previously found solution for the matter dominated universe, which we denote with an ``EdS" subscript.
The corresponding approximation for the one-loop power spectrum is
\bea
P(k,\tau)\amp=\amp D(\tau)^2P_{11}(k)\nonumber\\
\amp+\amp D(\tau)^4\big{[}2P_{13}(k)+P_{22}(k)\big{]}_{\mbox{\tiny{EdS}}}
\label{resumP}\eea
where $P_{11}(k)$ is the expression from eq.~(\ref{P11W}), $P_{13}(k)$ is the expression from (\ref{P13IR}), and $P_{22}(k)$ is the expression from (\ref{P22IR}).  
The reason this approximation works well is ultimately due to the fact that the dimensionless matter density $\Omega_m$ is approximately given by
\beq
\Omega_m\approx f^2
\eeq
in a $\Lambda$CDM universe at all times. A full treatment of the time dependence was performed recently in \cite{Carrasco:2012cv} in terms of numerically evaluated Green's functions. These results can be compared and are found to be remarkably similar. The approximate results obtained provide useful analytical results and intuition, while the full Green's functions can be used for improved accuracy.

As before, $P_{13}$ includes a UV contribution from the fluid parameters. Generalizing the previous result from (\ref{P13UV}) to the $\Lambda$CDM case, we have
\beq
P_{13,UV}(k,\Lambda)
= -{(c_{s,0}^2(\Lambda)+f_0\,c_{v,0}^2(\Lambda))k^2\over9f_0^2\mathcal{H}_0^2D_0^2}P_{11}(k)\label{P13UVGen}
\eeq
which is of the form reported earlier in the introduction in eq.~(\ref{Pcorr}) (where we suppressed the detailed time dependence).
The $\Lambda$ dependence of the fluid parameters is given by
\beq
c_{s,0}^2(\Lambda)+f_0\,c_{v,0}^2(\Lambda)
=\left({488\over5}{1\over504}{9\mathcal{H}_0^2f_0^2D_0^2\over 4\pi^2} \int_\Lambda dq\,P_{11}(q)\right)
+c_{s,0}^2(\infty)+f_0\,c_{v,0}^2(\infty)
\label{csLamDepLCDM}\eeq
Evidently, the important combination is $c_s^2+f c_v^2$, which is the value we measured and reported on in Fig.~\ref{csLambda}.

\subsection{Summing the Linear Terms}\label{Resum}

In the previous section we treated the fluid terms perturbatively. This meant  we took the leading term in the expansion to be the usual growing mode in a matter dominated era $\delta_1=f\,\theta_1\propto D(a)$, and the fluid parameters provided corrections in an expansion in powers of the scale factor. However, since the sound speed and viscosity enter the linear theory, we can resum their contributions and form a new type of expansion in powers of the density field. In this way, all linear terms enter at first order, and only non-linear terms enter at second order, etc. This method treats the sounds speed and viscosity as independent parameters that can be measured separately. Although this is not necessarily possible in practice due to degeneracy with higher order terms in the stress-tensor expansion, as we explain in Section \ref{Degeneracy}, this gives a sense of the consequences of summing a large number of terms.

\begin{figure}[t]
\center{\includegraphics[width=11cm]{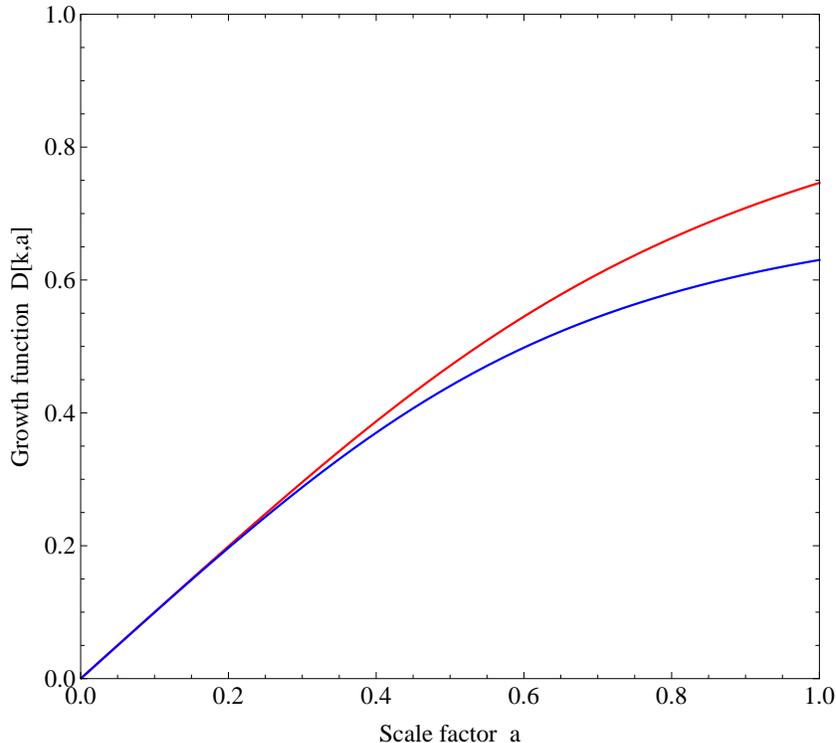}}
\caption{Re-summed growth function $D(k,a)$ with $k$ fixed at $k$ as a function of scale factor $a$, with standard cosmological parameters for a $\Lambda$CDM universe.
The upper (red) curve is for $k=0$, i.e, the usual growth function in a $\Lambda$CDM universe, which also coincides with SPT for all $k$ with vanishing fluid parameters. The lower (blue) curve is for the EFT with $k=0.2$\,[h/Mpc] and representative fluid parameters $c_s^2=7.2\times10^{-9}c^2$ and $c_v^2=2.7\times10^{-9}c^2$ (although they exhibit degeneracy, as we discuss in Section \ref{Degeneracy}).}
\label{FirstOrderNew}\end{figure}

Let us write the expansion schematically as
\bea
\delta_l({\bf k},\tau)\amp=\amp\sum_{n=1}^\infty \delta^{(n)}({\bf k},\tau)\label{deltaschem}\\
\theta_l({\bf k},\tau)\amp=\amp\sum_{n=1}^\infty \theta^{(n)}({\bf k},\tau)\label{thetaschem}
\eea
The equations of motion (\ref{Conttau},\ref{Eulertau}) give us the following linear order equations
\bea
&&{d\delta^{(1)}\over d\tau}+\theta^{(1)}= 0\label{ConttauLin}\\
&&{d\theta^{(1)}\over d\tau}+\mathcal{H}\theta^{(1)}+{3\over2}\mathcal{H}^2\Omega_m\delta^{(1)}
\nonumber\\ &&\,\,\,\,\,\,\,-c_s^2k^2\delta^{(1)}+{c_v^2k^2\over\mathcal{H}}\theta^{(1)}=0\label{EulertauLin}
\eea
By substituting eq.~(\ref{ConttauLin}) into eq.~(\ref{EulertauLin}) and re-arranging, we obtain a second order ODE for $\delta^{(1)}$. Lets express $\delta^{(1)}$ in terms of a growth factor $D(k,\tau)$ and a stochastic variable $\delta_1({\bf k})$, i.e.,
\beq
\delta^{(1)}({\bf k},\tau)=D(k,\tau) \delta_1({\bf k})\label{deltaD}
\eeq
The growth factor $D$ satisfies the same ODE as $\delta^{(1)}$, namely
\beq
{d^2 D\over d\tau^2}+\mathcal{H}\left(1+{c_v^2k^2\over\mathcal{H}^2}\right){d D\over d\tau}
-{3\over2}\mathcal{H}^2\left(\Omega_m-{2c_s^2k^2\over3\mathcal{H}^2}\right)D=0\label{ODE}
\eeq
We impose the asymptotic  condition $D(k,\tau)\to a(\tau)$ for small $a$. The solution is plotted in Figure \ref{FirstOrderNew} for typical values of $c_s^2,\,c_v^2$.

Although there is no simple analytical form for $D(k,\tau)$, we can exhibit its structure. In particular, it has a self similar behavior, making it (up to a rescaling) only a function of a combination of a particular product of $k,\,\tau$, rather than $k$ and $\tau$ independently (the product is $k\,\tau$ when the fluid parameters are treated as time independent and $\sqrt{k}\,\tau$ when they are treated as time dependent).
For the case in which we take $c_s^2,\,c_v^2$ to be time independent, let us rescale time to the following dimensionless variable
\beq
T\equiv \sqrt{c_s c_v} \, k \, \tau
\eeq
We then find that in a matter dominated universe the ODE (\ref{ODE}) simplifies to
\beq
{d^2D\over dT^2}+{2\over T}\left(1+b{T^2\over 4}\right){dD\over dT}-{6\over T^2}\left(1-{1\over b}{T^2\over6}\right)D=0
\label{ODET1}\eeq
where $b\equiv c_v/c_s$. Of course this ODE has infinitely many solutions. Let us focus on one particular solution, which we denote $\mathcal{D}(T)$, that satisfies the special asymptotic condition $\mathcal{D}(T)\to T^2$ for small $T$. 
For any value of $k$ the solution for $D(k,\tau)$ is obtained from the one parameter function $\mathcal{D}(T)$ by 
\beq
D(k,\tau)={\mathcal{D}(\sqrt{c_s c_v} \, k \, \tau )\over c_s c_v k^2\tau_0^2}
\eeq
This clearly has the correct asymptotic behavior $D(k,\tau)\to a=(\tau/\tau_0)^2$ for small $a$, since $\mathcal{D}(T)\to T^2$ for small $T$.

In the case in which we take the fluid parameters to be time dependent, as examined in Section \ref{Timedepfluid} with time dependence given in eq.~(\ref{csvtime}), the analysis is slightly altered. In this case we introduce the dimensionless variable
\beq
T\equiv{(c_{s,0}c_{v,0})^{1\over4}\sqrt{k}\over\sqrt{\tau_0}}\,\tau
\eeq
and the corresponding ODE in a matter dominated universe is
\beq
{d^2D\over dT^2}+{2\over T}\left(1+b_0{T^4\over 4}\right){dD\over dT}-{6\over T^2}\left(1-{1\over b_0}{T^4\over6}\right)D=0
\label{ODET2}\eeq
where $b_0\equiv c_{v,0}/c_{s,0}$.
Note the different powers of $T$ in the parenthesis, compared to eq.~(\ref{ODET1}).
Again we define the function $\mathcal{D}(T)$ as the solution to this ODE with asymptotic condition $\mathcal{D}(T)\to T^2$ for small $T$. The corresponding solution for the growth factor is 
\beq
D(k,\tau)={\mathcal{D}\!\left((c_{s,0}c_{v,0})^{1\over4}\sqrt{k}\,\tau/\sqrt{\tau_0}\right)\over \sqrt{c_{s,0}c_{v,0}}\,k\,\tau_0}
\eeq

For the power spectrum, we simply use the same form as before (\ref{resumP}), but now dropping the $F_{3,1}$ terms as they are built (and re-summed) into the linear piece.
This approximation for $P(k,\tau)$ is overly simplistic, however, since the growth function $D$ is $k$ dependent. A better approximation is to embed $D(k,\tau)$ inside the convolution integrals of (\ref{P13IR},\ref{P22IR}).
Indeed we expect it to give somewhat accurate results, as has been the case in related calculations \cite{Takahashi:2008yk} and we shall use this approximation in Section \ref{PowerNumResults}. The differential equations, whose solutions give the first few terms in the exact expansion, are provided in Appendix \ref{ExactExp}.

\section{Power Spectrum Results}\label{PowerNumResults}

By matching to N-body simulations, as described in detail in Ref.~\cite{Carrasco:2012cv}, one can measure the linear combination
$c_{s,0}^2+f_0\,c_{v,0}^2$. For simulation parameters: $\Omega_m= 0.25$, $\Omega_\Lambda = 0.75$, $h = 0.7$  ($H = 70~{\rm [km/s/Mpc]}$), $\sigma_8 = 0.8$, and $n_s = 1$, with measurements described taking place at $z=0$, and choosing a smoothing scale of $\Lambda=1/3$ [h/Mpc] it was found
\beq
c_{s,0}^2+f_0\,c_{v,0}^2\approx 9\times 10^{-7}\,c^2.
\eeq

Having obtained this linear combination, this completes the required quantities in order to compute the one-loop power spectrum that we derived earlier to the desired approximation.
The single combination $c_s^2+f c_v^2$ can be used as a single insertion by the formulae for $P(k)$ derived in Section \ref{PSpectrum},
or we can assume approximate individual values for $c_s^2$ and $c_v^2$ separately in the re-summed formulae for the growth function $D(k,\tau)$ and hence $P(k)$ as indicated in Section \ref{Resum}, although there is degeneracy in their values as we explain later in Section \ref{Degeneracy}.

\begin{figure}[t]
\center{\includegraphics[width=11.5cm]{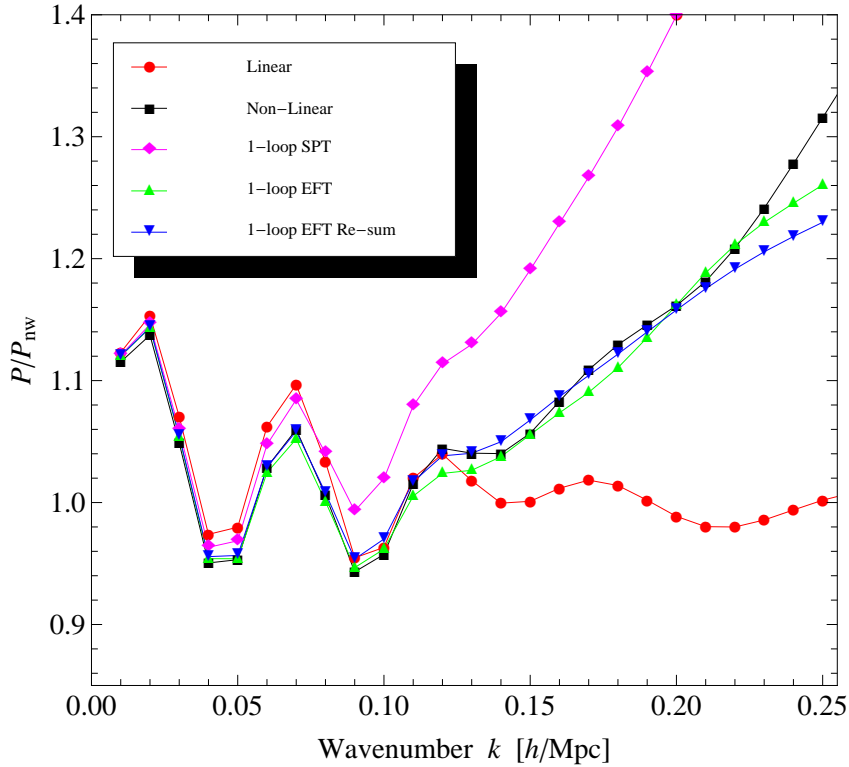}}
\caption{Power spectrum, normalized to the no-wiggle spectrum of \cite{Eisenstein:1997ik} at $z=0$ for $\Lambda$CDM universe.
Red is the linear theory (from CAMB), black is the full non-linear reference value, magenta is the one-loop SPT, green is the one-loop EFT with a single insertion of fluid parameters, and blue is the one-loop EFT with a re-summation of fluid parameters as discussed in Section \ref{Resum}.}
\label{PowerOneLoop2}\end{figure}

Our results for the full power spectrum $P(k)$ in both of these approximations is given in Fig.~\ref{PowerOneLoop2}. It is normalized to the no-wiggle power spectrum $P_{nw}(k)$ of \cite{Eisenstein:1997ik} is the linear power spectrum without baryon-acoustic-oscillations.
For convenience, we have taken the large $\Lambda$ limit, by using the $\Lambda$ dependence that we derived earlier in eq.~(\ref{csLamDepLCDM}).
This causes the fluid parameters to asymptote to a slightly lower value, roughly
\beq
c_{s,0}^2(\infty)+f_0\, c_{v,0}^2(\infty) \approx 8\times 10^{-7}\,c^2
\eeq
In Fig.~\ref{PowerOneLoop2} we have also included the result for SPT for comparison and a non-linear reference value.
We see that the power spectra of the EFT are much better than both the linear theory and SPT. The re-summed case is arguably better at lower $k$, though it is somewhat degenerate with higher order effects, and since the other higher order effects are not included here there is some disagreement at higher $k$. The single insertion is  very accurate also. The fact that the fluid parameters asymptote to a finite non-zero value in the $\Lambda\to\infty$ limit represents the finite error made in SPT. When made dimensionless, by multiplicity by $c^2k^2/H^2$ (say at $z=0$), it leads to an important correction to the power spectrum as we increase $k$, as seen in Fig.~\ref{PowerOneLoop2}.
Note that the EFT result is quite accurate; roughly at the 1\% level out to $k\sim\mbox{few}\times 0.1$\,h/Mpc.

\section{Discussion}\label{Discussion}

In this section we briefly mention some interesting issues surrounding the effective fluid, including its Reynolds number, degeneracy of parameters, the inclusion of collisions or wave-like behavior, higher order moments, and the velocity field.

\subsection{Reynolds number}
For viscous fluids there is a famous dimensionless number which captures its tendency for laminar or turbulent flow; the ``Reynolds number".
The Reynolds number is defined as
\beq
R_e\equiv {\rho \, v L\over\eta}
\eeq
where $\eta$ is shear viscosity, $\rho$ is density, $v$ is a characteristic velocity, and $L$ is a characteristic length scale. This is
\beq
R_e\sim {H v L\over c_{sv}^2}\sim {H^2a^2\over c_{sv}^2 k^2}\delta\lesssim 10
\eeq
Hence the Reynolds number is not very large, and the system is therefore not turbulent.
Furthermore, if we were to estimate the viscosity by Hubble friction, then we would have $R_e\sim\delta$ and so the Reynolds number would be even smaller in the linear or weakly non-linear regime.

For cosmological parameters $\rho_b\sim 3\times 10^{-30}$\,[g/cm$^3$], $H=70$\,[km/s/Mpc],
and if we take a plausible value for the shear viscosity of $c_{sv}^2\sim 2\times 10^{-7}c^2$, then the viscosity coefficient
is found to be 
\beq
\eta\sim 20\, \mbox{Pa\,s}
\eeq
which is perhaps surprisingly not too far from unity in SI units. (For instance, it is somewhat similar to the viscosity of some everyday items, such as  chocolate syrup.) A proper measurement of $c_{sv}^2$ would come from a detailed measurement of vorticity; a point that we will return to in the following subsection.

\subsection{Degeneracy in Parameters}\label{Degeneracy}

Earlier we discussed the individual parameters: the sound speed $c_s^2$, shear viscosity $c_{sv}^2$, and bulk viscosity $c_{bv}^2$. We showed that by taking the curl of the Euler equation we obtained an equation for vorticity that involves $c_{sv}^2$. Hence a careful analysis of vorticity could reveal the value of $c_{sv}^2$ - such a value would enter our discussion of ``Reynolds number" of the previous subsection, although not the bulk of this paper. Since the vorticity is rather small, this would be non-trivial to measure, though possible.

On the other hand, by taking the divergence of the Euler equation we obtained coupled equations for $\theta_l$ and $\delta_l$ that involves the sound speed $c_s^2$ and the combination of viscosity $c_v^2=c_{sv}^2+c_{bv}^2$. One could try to measure these two parameters independently using eqs.~(\ref{csx},\,\ref{cvisx}) from N-body simulations. However one should be careful as to how to interpret this result. At the one-loop level, we saw that it was only a certain linear combination that appeared in the result, namely $c_s^2+f\,c_v^2$. In other words, the two parameters appear in a degenerate way at one-loop. This degeneracy would be broken at higher loop order. However, to be self-consistent one should then also include new couplings (for instance, representing higher derivative operators in the stress-tensor expansion) which would also enter, leading to a new constraint and a new type of degeneracy with the new parameters.

This reason for this degeneracy is the following: In a universe in which one could  track the full evolution of the initial state, one would observe that $c_s^2$ and $c_v^2$ affect the one-loop theory differently. However, in the real universe, there is a growing mode and a decaying mode. In practice, one does not track the decaying mode, only the growing mode, as studied here in this paper. For this single mode the parameters enter in a special linear combination.

\subsection{Interactions}\label{Collisions}
In this paper, we have treated dark matter as being comprised of collisionless particles, interacting only through gravity. Of course we expect that there are also some finite non-gravitational interactions. These effects can be treated perturbatively in the effective field theory by identifying the relevant length scale, which is the mean free path between scattering. 

As an example, for WIMPs the scattering cross section is roughly $\sigma\sim g^2/m_W^2$,
where $g\sim 0.1$ is a dimensionless coupling and $m_W\sim 100$\,GeV is of order the weak scale.
The mean free path between scatterings is
\beq
\lmfp={1\over n\,\sigma}\sim {m_W^3\over\rho_b\, g^2}
\eeq
The background density, in Planck units, is $\rho_b\sim 10^{-120}\mpl^4$. Hence,
\beq
\lmfp\sim 10^{17}d_H\left(0.1\over g\right)^2\left(m_W\over100\,\mbox{GeV}\right)^3
\eeq
where $d_H=1/H_0$ is the present Hubble length.
As another example, for QCD-axions scattering due the $~\phi^4$ term in the potential $V(\phi)=\Lambda_{bcd}^4(1-\cos(\phi/F_{PQ}))$, the mean free path is much larger still.

On the other hand, gravity introduces a non-linear scale of the order $\lnl\sim 10^{-4}\,d_H$ (which can be thought of as the mean free path between gravitational scattering of a point particle off a large non-linear clump). Since the collisional mean free path in these examples satisfies $\lmfp\gg\lnl$ it can be ignored at first approximation. Though in principle it can be included perturbatively in the effective field theory, but suppressed by a hierarchy $\lnl/\lmfp$, which may be of interest for some highly non-standard dark matter candidates.

\subsection{Wave-like Behavior}\label{Wave}

In this paper we have treated the dark matter as comprised of classical point-like particles. 
This obviously ignores its quantum mechanical wave-like behavior. For most dark matter candidates, such as a typical WIMP with a weak scale mass, the de Broglie wavelength is extremely small and ignorable.
For extremely light (pseudo)-scalars, such as axions, it is conceivable that their de Broglie wavelength is large and relevant.

For a classical scalar field and also for a Bose-Einstein condensate \cite{Sikivie:2009qn}, one can show that in the linear theory, there is a correction to the pressure of the form
\beq
\delta p=-{\hbar^2\over 4\,a^2 m^2}\nabla^2\rho
\eeq
this provide a contribution to a type of scale dependent sound speed $\delta c_s^2\sim \hbar^2 k^2/(a^2m^2)$.
Now recall that the characteristic correction from the sound speed is $\sim c_s^2k^2/(H^2a^2)$. This means that a rough estimate for the dimensionless correction from the quantum character of the particles is (at $z=0$)
\beq
\mbox{quantum correction} \sim {\hbar^2 k^4\over m^2H_0^2}
\eeq
The relative size of this contribution obviously depends on the mass of the particle $m$.
It is important to note that in the point-particle treatment, the mass $m$ dropped out of all results. But by including such UV physics, we gain more sensitivity in the effective field theory to such physical parameters.

The dark matter particle mass $m$ can in principle be very small. For instance, in the so-called string axiverse it is suggested that there may be a range of extremely light axions \cite{Arvanitaki:2009fg}; one of which could provide the bulk of the dark matter (though there are important constraints from isocurvature bounds on light axions \cite{Hertzberg:2008wr,Hamann:2009yf}, while the classic axion window is still very promising \cite{Hertzberg:2012zc,Erken:2011dz}). Here we would like to mention that the mass of a dark matter particle presumably cannot be arbitrarily small because its de Broglie wavelength $\lambda_{dB}\sim \hbar/(m v)$ would then smear it out over scales larger than that of a 
galaxy $L_{gal}$, and yet we know dark matter clumps on galactic scales. By imposing $\lambda_{dB}<L_{gal}$ this gives the bound
\beq
m>{\hbar\over L_{gal}v}
\eeq
Hence we have a bound on the dimensionless correction from the wave-like character of light scalars as
\beq
\mbox{quantum correction}<{k^4L_{gal}^2v^2\over H_0^2}
\eeq
where $v$ is a characteristic dispersion velocity associated with the dark matter.
By estimating $v\sim 10^{-3}\,c$, then we can estimate $H_0/v\sim k_{NL}$, leading to the rough bound
\beq
\mbox{quantum correction}<{k^4L_{gal}^2\over k_{NL}^2}
\eeq
Since $L_{gal}$ is much smaller than the non-linear scale (for instance, $L_{gal}$ may be as small as dwarf galaxy size) we see that the quantum correction must always be very small in the regime in which the effective field theory is valid (i.e., $k<k_{NL}$).

\subsection{Higher Order Moments}

In principle, one can study higher order moment of the Boltzmann equation. In Section \ref{EffectiveFluid} we considered the zeroth moment (continuity) and first moment (Euler), and then built a derivative expansion for the effective stress-tensor that appears on the right hand side of the Euler equation.
The stress-tensor involves two contributions: kinetic and gravitational. The kinetic piece $\kappa_l^{ij}$ is includes the second moment of the velocity distribution (minus the long modes), and so it evolves under the second moment of the Boltzmann equation. The trace of the kinetic part of the stress-tensor is proportional to a type of ``kinetic temperature" $T\sim\kappa_l/\rho_l$. Although the system is not in thermal equilibrium, so this name is only by analogy to classical systems which are.

However, since the stress-tensor also includes the gravitational piece $w^{ij}$, we do not have an evolution equation for the full stress-tensor. Parametrically these two contributions are of the same order (for instance they cancel each other in the virial limit). So this requires the use of a derivative expansion to capture the effects of these higher order moments and interactions, in the effective field theory sense.

\subsection{Velocity Field}

Let us also make some comments on the computation of correlation functions, such as the two-point, involving the smoothed velocity field.
In Section \ref{CutoffIndependence} we demonstrated how the cutoff dependence in the expansion for $\delta_l$ cancels out when we form the two-point correlation function for density, which involved the fluid parameter's canceling the $\Lambda$ dependence of the $P_{13}$ loop. However, once the fluid counter-terms are introduced to cancel this dependence, then they cannot also be used to cancel the cutoff dependence appearing in the loops of correlation functions of other fields.
In particular, consider the velocity field. Recall that it is defined by the ratio $v^i_l=\pi_l^i/\rho_l$. If we Fourier transform this, then examine the regime $k<\Lambda$, we are still left with $\Lambda$-dependence, even though both $\rho_l$ and $\pi^i_l$ are cutoff independent for $k<\Lambda$. 
This is similar to certain kinds of non-linear objects that one might define to describe pions, such as a bi-linear in the quark fields, which depends explicitly on the cutoff.
This means that the velocity field $v_l^i$ is inherently cutoff dependent even in the $k\ll\Lambda$ regime, while $\delta_l$ is not.

\section{Summary and Outlook}\label{Summary}

In this paper we have examined and developed the effective field theory of dark matter and structure formation on sub-horizon scales, emphasizing detailed analytical and semi-analytical results, including a recursion relation for the perturbative expansion and an approximate extraction of the time dependence of the growing modes. This works compliments and extends the important recent work in Ref.~\cite{Carrasco:2012cv}. 
These works can be viewed as a precise realization of the conceptual foundation laid out in Ref.~\cite{Baumann:2010tm}, where special focus was placed on the issue of back-reaction at the scale of the horizon, though the present focus is on sub-horizon scales and the explicit computation of the power spectrum.  The effective field theory is an expansion for wave numbers $k$ less than the cosmological non-linear scale $\knl$. It is a cosmological fluid description for cold dark matter, and by extension all matter including baryons which trace the dark matter.
The microphysical description was in terms of a classical gas of point particles, which we smoothed at the level of the Boltzmann equation and used the Newtonian approximation for sub-horizon modes.
We exhibited the various couplings that appear in the effective field theory, namely pressure and viscosity, whose linear combination was obtained  by matching to N-body simulations in Ref.~\cite{Carrasco:2012cv}, finding $c_s^2+f\,c_v^2\sim 10^{-6}\,c^2$. 
This represents the finite error made in standard perturbation theory and has important consequences for the power spectrum.
We see that standard perturbation theory would only be correct if the linear power spectrum was very UV soft so that the theory remains perturbative to arbitrarily high $k$; this would allow one to send $\Lambda\to\infty$ and there would be no stress-tensor at all as it is sourced only be the short modes. However the presence of the non-linear scale forces one to introduce the cutoff $\Lambda$ and a finite stress-tensor, which evidently does not vanish in the large $\Lambda$ limit (which can be formally taken order by order after the $\Lambda$ dependence is cancelled).

We developed the perturbative expansion for the power spectrum, which we recast into a recursive formula and extracted the time dependence in a convenient way. The power spectrum was then computed at the one-loop order. 
We found that the corrections from the fluid parameters led to a power spectrum in good agreement with the full non-linear spectrum. 
Unlike the standard perturbation theory that deviates substantially from the true non-linear power spectrum, especially at low $z$, suggesting
that the non-linear wavenumber is low, the effective field theory exhibits $\sim 1\%$ level accuracy for $k\sim\mbox{few}\times 0.1$\,h/Mpc, suggesting that the true non-linear wavenumber may be higher than ordinarily thought. Furthermore, the general success of the effective field theory approach suggests that any deviations from the $\Lambda$CDM model should fit into this framework by altering the fluid parameters.

The effective field theory approach to large scale structure formation is complimentary to N-body simulations by providing an elegant fluid description. 
This  provides intuition for various non-linear effects, as well as providing computational efficiency, since the numerics required to measure the fluid parameters can be less computationally expensive than a full scale simulation. Of course, since the couplings are  UV sensitive, it still requires the use of some form of N-body simulation to fix the physical parameters, either by matching to the stress-tensor directly or to observables; a point analyzed in detail in Ref.~\cite{Carrasco:2012cv}. But this matching is only for a small number of physical parameters at some scale and then the constructed field theory is predictive at other scales.
The formulae for the power spectrum, exhibited in eqs.~(\ref{Pexp}--\ref{P13UV},\,\ref{csLamDep}) for matter dominated and eqs.~(\ref{resumP}--\ref{csLamDepLCDM}) for $\Lambda$CDM, should be quite powerful and convenient in this area of cosmological research.

There are several possible extensions of this work. A first extension is to go beyond the one-loop order to two-loop, or higher. 
It is important to note that the treatment can in principle capture the full power spectrum to arbitrary accuracy if carried out to the required order. Here we have computed the power spectrum at one-loop, which is order $\delta^4$, but higher order is possible.
This will require the measurement of several new parameters that will enter the effective stress-tensor at higher order. Another extension is to fully measure the stochastic fluctuations, which will involve measuring the correlation function of the stress-tensor with itself. These effects are somewhat reduced at low $k$, especially due to the suppression of modes in the integrand due to the turnover in the transfer function in eq.~(\ref{P22UV}), but are of significant interest and should improve the agreement at higher $k$.
Another extension is to include the small but finite contributions from vorticity, or to compute higher order N-point functions, which can probe non-Gaussianity, or to consider different cosmologies, and to include baryons, etc.

In general it is essential to gain insight and precision into the mapping between the microphysics that determines the early universe, including the distribution of primordial fluctuations and the contents of the universe, and the output universe that we can observe today. This approach, when complimented with N-body simulations and observations, may ultimately give new insights into fundamental questions in cosmology.
This is an exciting and promising way to learn about fundamental physics.

\section*{Acknowledgments}
We would like to thank Tom Abel, Roger Blandford, John Joseph Carrasco, Leonardo Senatore, and Risa Weschler for helpful discussions. MH is supported by SITP, KIPAC, NSF grant PHY-0756174, and a Kavli Fellowship.

\appendix

\section{Short Modes}\label{Shortmodes}

Although we use the full stress-tensor in (\ref{stressT})--(\ref{gravT}) as a generating functional of the effective theory, in this appendix we demonstrate that we can separate out the long modes from the short modes in the Euler equation. To do so, we define the short modes to be
\bea
\sigma_s^{ij}\amp\equiv\amp m^{-1}a^{-5}\!\int\! d^3{\bf p}\,(p^i-p^i_l({\bf x}))(p^j-p^j_l({\bf x}))f({\bf x},{\bf p}) \\
\amp=\amp \sum_n {m\over a^3}(v_n^i-v_l^i({\bf x}_n))(v_n^j-v_l^j({\bf x}_n))\,\delta^3_D({\bf x}-{\bf x}_n) \,\,\,\,\,\,\,\,\,\,\, \\
\phi_{s,n}\amp\equiv\amp \phi_n-\phi_{l,n}\\
\partial_i\phi_s\amp=\amp \sum_n\partial_i\phi_{s,n}\\
w^{ij}_s \amp\equiv\amp 
\partial_{i}\phi_s\,\partial_{j}\phi_s-\sum_n\partial_i\phi_{s,n}\,\partial_j\phi_{s,n}
\eea
where $p_l^i({\bf x})\equiv m \, a \, v_l^i({\bf x})$.
Note that $\sigma^{ij}_s\neq\sigma^{ij}-\sigma_l^{ij}$, but they are related as follows
\bea
\sigma_l^{ij}\amp=\amp\left[\sigma_s^{ij}\right]_\Lambda+\left[\rho_mv_l^iv_l^j\right]_\Lambda+
\left[v_l^i(\pi^j-\rho_mv_l^j)+v_l^j(\pi^i-\rho_mv_l^i)\right]_\Lambda
\eea
The second term is approximately $\rho_lv_l^iv_l^j$ (so it approximately cancels with $-\rho_lv_l^iv_l^j$ in $\kappa_l^{ij}$) and the final term is small (as it is an overlap between short and long modes). 
Following the methods of \cite{Baumann:2010tm} we obtain
\bea
\kappa_l^{ij}=\left[\sigma_s^{ij}\right]_\Lambda+{\rho_l\partial_kv_l^i\partial_kv_l^j\over\Lambda^2}+\mathcal{O}\left(1\over\Lambda^4\right)
\eea
Similarly, one can prove that $\Phi^{ij}_l$ satisfies
\bea
\Phi^{ij}_l \amp=\amp -{[w^{kk}_s]_\Lambda\delta^{ij}-2[w^{ij}_s]_\Lambda\over 8\pi G\,a^2}+\nonumber\\
\amp\amp \!\!\!\!\!{\partial_m\partial_k\phi_l\partial_m\partial_k\phi_l\delta^{ij}-2\partial_m\partial_i\phi_l\partial_m\partial_j\phi_l\over 8\pi G\,a^2\Lambda^2}+\mathcal{O}\!\left(1\over\Lambda^4\right)\,\,\,\,\,\,\,\,\,\,
\eea

So altogether we obtain the effective stress-tensor
\bea
\left[\tau^{ij}\right]_\Lambda \amp=\amp\left[\tau_s^{ij}\right]_\Lambda+\left[\tau^{ij}\right]^{\partial^2}
\eea
where
\bea
\left[\tau_s^{ij}\right]_\Lambda\amp=\amp \left[\sigma_s^{ij}\right]_\Lambda
-{[w^{kk}_s]_\Lambda\delta^{ij}-2[w^{ij}_s]_\Lambda\over 8\pi G\,a^2}  \\
\left[\tau^{ij}\right]^{\partial^2}\amp=\amp {\rho_l\partial_kv_l^i\partial_kv_l^j\over\Lambda^2}+
{\partial_m\partial_k\phi_l\partial_m\partial_k\phi_l\delta^{ij}
-2\partial_m\partial_i\phi_l\partial_m\partial_j\phi_l\over 8\pi G\,a^2\Lambda^2}+\mathcal{O}\!\left(1\over\Lambda^4\right)\,\,\,\,\,\,\,\,\,\,
\eea
We see that $[\tau^{ij}]_\Lambda$ is sourced by short wavelength fluctuations plus higher derivative corrections.

Note that by taking the derivative $\partial_j$ this leading piece becomes
\bea
\partial_j[\tau^{ij}_s]_\Lambda=\partial_j\!\left[\sigma_s^{ij}\right]_\Lambda+\left[\rho_s\partial_i\phi_s\right]_\Lambda
\eea
with
\beq
[\rho_s\partial_i\phi_s]_\Lambda=\sum_{n\neq\bar{n}}m\,a^{-3}\partial_i\phi_{s,\bar{n}}({\bf x}_{n})W_\Lambda({\bf x}-{\bf x}_{n})-[\rho_l\partial_i\phi_s]_\Lambda
\label{shortgrav}\eeq
where the first term in (\ref{shortgrav}) is given by
\bea
&&\sum_{n\neq\bar{n}}m\,a^{-3}\partial_i\phi_{s,\bar{n}}({\bf x}_{n})W_\Lambda({\bf x}-{\bf x}_{n})\nonumber\\
&&=\sum_{n\neq\bar{n}}{m^2G\over a^4}{(x_n-x_{\bar{n}})^i \over |{\bf x}_n-{\bf x}_{\bar{n}}|^3}\left(
\mbox{Erfc}\left[{\Lambda|{\bf x}_n-{\bf x}_{\bar{n}}|\over\sqrt{2}}\right]+{4\pi|{\bf x}_n-{\bf x}_{\bar{n}}|\over\Lambda^2}W_\Lambda({\bf x}_n-{\bf x}_{\bar{n}})\right)W_\Lambda({\bf x}-{\bf x}_n)\,\,\,\,\,\,\,\,\,\,\,\,\,\,\,
\eea
and the second term in (\ref{shortgrav}) can be expanded as
\beq
[\rho_l\partial_i\phi_s]_\Lambda=-{1\over 2\Lambda^2}\rho_l\partial_i\partial^2\phi_l+\ldots
\eeq
and this term should be included since it involves the background piece $\rho_b$, and so it includes a first order contribution.

\section{Trace of Stress-Tensor}\label{TraceTensor}

It is of some interest to compute the trace of the stress-tensor, which is the so-called ``mechanical pressure". This includes the gravitational piece
\bea
\Phi_l\amp=\amp-{w^{kk}_l\over 8\pi G\,a^2}+{\partial_k\phi_l\partial_k\phi_l\over 8\pi G\,a^2}
\eea
The first term is approximately given by
\bea
-{w^{kk}_l\over8\pi G\,a^2} 
\amp\approx\amp{1\over2}\int d^3{\bf x}'\,W_\Lambda({\bf x}-{\bf x}')
\Big{[}\delta\rho({\bf x}')\phi({\bf x}')-\sum_n \delta\rho_n({\bf x}')\phi_n({\bf x}') \Big{]}\label{approxtrace}\\
\amp=\amp-{1\over2}\sum_{n\neq\bar{n}}{Gm^2\over a^4|{\bf x}_n-{\bf x}_{\bar{n}}|}e^{-\mu|{\bf x}_n-{\bf x}_{\bar{n}}|}W_\Lambda({\bf x}-{\bf x}_n)\nonumber\\
\amp\amp
+{1\over 2}\sum_n{4\pi Gm\rho_b\over a\mu^2}W_\Lambda({\bf x}-{\bf x}_n)
\eea
where we have used the identity
\beq
(\nabla\phi)^2=-\phi\,\nabla^2\phi+{1\over2}\nabla^2(\phi^2)
\eeq
and dropped all terms that are suppressed by the ratio of low $k$-modes to high $k$-modes in (\ref{approxtrace}). 
If we let an $s$-subscript denote the short modes (see Appendix \ref{Shortmodes} for details), then the trace of the stress-tensor is roughly
\bea
[\tau]_\Lambda\approx \int d^3{\bf x}'\,W_\Lambda({\bf x}-{\bf x}')\!\!\!\!\!\amp\amp\Big{[}\rho({\bf x}')\!\Big{(}v_s({\bf x}')^2+{1\over 2}\phi_s({\bf x}')\Big{)}-{1\over2}\sum_n \rho_{s,n}({\bf x}')\phi_n({\bf x}') \Big{]}
\eea
The background pressure has the zero mode contribution
\beq
p_b={1\over 3}\langle[\tau]_\Lambda\rangle
\label{bkgdpressure}\eeq
where we have ignored a correction from the bulk viscosity. There are also stochastic fluctuations to the pressure, which we discuss separately in Appendix \ref{StochasticFluctuations}.
Now, since the density field $\rho({\bf x})$ can be arbitrarily large for dense objects on small scales, it suggests that each of the contributions to the renormalized pressure, both the kinetic and the gravitational, can be quite large. 
However, for virialized structures, these two terms cancel each other \cite{Baumann:2010tm}. 
Hence the only significant contribution to the integral comes from modes of order $k\sim \knl$ which have yet to virialize. For a pure Einstein de Sitter universe, this leads to the estimate $p_b\sim 10^{-5}\rho_bc^2$, where the factor of $10^{-5}c^2$ is the typical value of the potential $\phi$ from the primordial power spectrum. 
However, due to the turn-over in the power spectrum at matter-radiation equality, we are led to an value that is about an order of magnitude smaller; see Section \ref{Timedepfluid}.

\section{Exact Expansion}\label{ExactExp}

The expansion in Section \ref{Resum} should be compared to the exact expansion, which we briefly mention. The second order density fluctuation $\delta^{(2)}$ can be written in terms of a pair of time dependent kernels $D_{2A}$ and $D_{2B}$ as follows
\bea
\delta^{(2)}({\bf k},\tau)\amp=\amp{1\over7}\int\!{d^3k'\over(2\pi)^3}\delta_1({\bf k}-{\bf k}')\delta_1({\bf k}')\nonumber\\
\amp\amp\,\,\,\,\,\times\Big{[}5\alpha({\bf k},{\bf k}')D_{2A}({\bf k},{\bf k}',\tau)
+2\beta({\bf k},{\bf k}')D_{2B}({\bf k},{\bf k}',\tau)\Big{]}
\label{delta2exact}\eea
The kernels satisfy the following ODEs
\bea
\hat{L}\, D_{2A}\amp=\amp{7\over5}\Big{[}{dD\over d\tau}(|{\bf k}-{\bf k}'|){d D\over d\tau}(k')
+{3\over2}\mathcal{H}^2\Big{(}\Omega_m-{2c_s^2k^2\over3\mathcal{H}^2}\Big{)}D(|{\bf k}-{\bf k}'|)D(k')\Big{]}\,\,\,\,\,\,\,\,\,\,\,\label{D2A}\\
\hat{L}\, D_{2B}\amp=\amp{7\over2}{dD\over d\tau}(|{\bf k}-{\bf k}'|){dD\over d\tau}(k')\label{D2B}
\eea
where $\hat{L}$ is the linear operator
\beq
\hat{L}\equiv{d^2\over d\tau^2}+\mathcal{H}\left(1+{c_v^2k^2\over\mathcal{H}^2}\right){d\over d\tau}-{3\over2}\mathcal{H}^2
\left(\Omega_m-{2c_s^2k^2\over3\mathcal{H}^2}\right)
\eeq
We have suppressed the $\tau$ dependence in the argument of $D$ and the $\tau,\, {\bf k},\, {\bf k}'$ dependence in the argument of $D_{2*}$ in (\ref{D2A},\,\ref{D2B}). 
As usual we impose the asymptotic condition $D_{2*}\to a^2$ for small $a$.
By numerically solving this pair of ODEs, we can compare to the approximation in (\ref{resum},\,\ref{resumP}) in which we replace $D_{2*}({\bf k},{\bf k}',\tau)\to D(k,\tau)^2$ in the power spectrum. 
At third order, we again expand $\delta^{(3)}$ in kernels $D_{3*}$ to obtain ODEs whose solutions can be compared to the approximation in (\ref{resum},\,\ref{resumP}) in which we replace $D_{3*}({\bf k},{\bf k}',{\bf k}'',\tau)\to D(k,\tau)^3$. These expressions can also be given in terms of Green's functions as we emphasized recently in \cite{Carrasco:2012cv}. As can be checked, the approximate and intuitive analytical results are quite close to the exact results.

\section{Stochastic Fluctuations and Pressure}\label{StochasticFluctuations}

The effective stress-tensor $[\tau^{ij}]_\Lambda$ fluctuates with the short modes, leading to a stochastic departure from the results computed thus far. This effect should be reduced at low $k$ since the transfer function softens the UV modes in the corresponding loop integral, but we would like to mention the formal procedure to include such finite size effects here, although we will not compute it precisely.

Again ignoring vorticity, but allowing for stochastic fluctuations, the evolution equations for the pair $\delta_l$ and $\theta_l$ are as given in eqs.~(\ref{Conttau},\,\ref{Eulertau}), but with an additional stochastic source term $j_s$
\bea
{d\delta_l\over d\tau}+\theta_l \amp=\amp -\int\!{d^3k'\over(2\pi)^3}\alpha({\bf k},{\bf k}')\delta_l({\bf k}-{\bf k}')\theta_l({\bf k}')\,\,\,\,\,\,\label{ConttauS}\\
{d\theta_l\over d\tau}+\mathcal{H}\theta_l+{3\over2}\mathcal{H}^2\Omega_m\delta_l
\amp=\amp-\int\!{d^3k'\over(2\pi)^3}\beta({\bf k},{\bf k}')\theta_l({\bf k}-{\bf k}')\theta_l({\bf k}')
+c_s^2k^2\delta_l-{c_v^2k^2\over\mathcal{H}}\theta_l    - j_s\,\,\,\,\,\,\,\,\,\,\label{EulertauS}
\eea
where $j_s=j_s({\bf k},\tau)$. In position space $j_s$ is defined from a scalar contraction of the fluctuations in the stress-tensor, namely
\beq
j_s\equiv {1\over \rho_b}\partial_i\partial_j\,\Delta\tau^{ij}
\eeq
where $\Delta\tau^{ij}$ is implicitly defined through eq.~(\ref{Ansatz}). So (up to higher order corrections) it is therefore related to the function $A_l$ by
\beq
a^2 A_l=c_s^2\partial^2\delta_l-{c_s^2 \partial^2\theta_l\over\mathcal{H}}+j_s
\eeq
Note that $j_s$ does not enter the two-point correlation functions we defined earlier, such as $\langle A_l\, \delta_l\rangle$, since the averaging annihilates any overlap between the non-stochastic pieces $\delta_l,\,\theta_l$ and the stochastic piece $j_s$. However, a two-point correlation function involving only the stochastic piece $\langle j_s\,j_s\rangle$ would be non-zero.

We treat the $j_s$ term as entering at higher order in a field expansion of the form (\ref{deltaschem},\,\ref{thetaschem}). As before, the first order term $\delta^{(1)}$ is given by the growth function $D$, defined through eqs.~(\ref{deltaD},\,\ref{ODE}). At second order, let us split the density field into two pieces
\beq
\delta^{(2)}=\delta^{(2)}_0+\Delta\delta^{(2)}
\label{delta2pieces}\eeq
where $\delta^{(2)}_0$ is defined as the contribution that arises even in the $j_s\to 0$ limit that we computed previously, i.e., $\delta_0^{(2)}$ is given by eq.~(\ref{delta2exact}) (or earlier approximate forms). While $\Delta\delta^{(2)}$ is the correction that arises from the stochastic contribution $j_s$. This new piece satisfies
\beq
\hat{L}\,\Delta\delta^{(2)}=j_s
\eeq
Formally, this may be solved by Green's functions. So let us introduce $G(t)$, defined as a particular solution of
the same differential equation, but with a Dirac-delta function RHS
\beq
\hat{L}\,G(k,\tau,\tau')=\delta^1_D(\tau-\tau')
\eeq
Then the contribution to the second order density from the stochastic fluctuations may be formally given by
\beq
\Delta\delta^{(2)}({\bf k},\tau)=\int\! d\tau'\,G(k,\tau,\tau')\,j_s({\bf k},\tau')
\eeq
We note that one can use this Green's function to express the solution in (\ref{delta2exact}).

The two-point correlation function $\langle\delta^{(2)}\,\delta^{(2)}\rangle$ has three contributions when we expand using (\ref{delta2pieces}). The first term $\langle\delta_0^{(2)}\,\delta_0^{(2)}\rangle$ is as we computed earlier.
The cross term $\langle\delta_0^{(2)}\,\Delta\delta^{(2)}\rangle$ vanishes, leaving only a single new term $\langle\Delta\delta^{(2)}\,\Delta\delta^{(2)}\rangle$, i.e.,
\beq
\langle\delta^{(2)}\,\delta^{(2)}\rangle=\langle\delta_0^{(2)}\,\delta_0^{(2)}\rangle+\langle\Delta\delta^{(2)}\,\Delta\delta^{(2)}\rangle
\eeq
The new term may be expressed in terms of the Green's functions and the stochastic contraction of the stress-tensor $j_s$ as
\bea
\amp\amp\langle\Delta\delta^{(2)}({\bf k},\tau)\,\Delta\delta^{(2)}({\bf k}',\tau)\rangle\nonumber\\
\amp\amp=\int\! d\tau'\!\int\! d\tau''\,G(k,\tau,\tau')\,G(k',\tau,\tau'')
\,\langle j_s({\bf k},\tau') j_s({\bf k}',\tau'')\rangle
\eea
This requires one to obtain the ensemble average $\langle j_s({\bf k},\tau') j_s({\bf k}',\tau'')\rangle$, which may be difficult to obtain numerically. Using isotropy we may partially simplify it to
\beq
\langle j_s({\bf k},\tau') j_s({\bf k}',\tau'')\rangle=(2\pi)^3\delta^3_D({\bf k}+{\bf k}')P_{jj}(k,\tau',\tau'')
\eeq
This leads to the following contribution to the two-point power spectrum
\bea
\Delta P_{22}(k,\tau)
\amp=\amp \int\! d\tau'\!\int\! d\tau''\,G(k,\tau,\tau')\,G(k,\tau,\tau'')
\,P_{jj}(k,\tau',\tau'')
\eea

This term should be understood as providing the UV part ($k>\Lambda$) of the $P_{22}$ Feynman diagram drawn in Fig.~\ref{FeynmanDiagrams}.
We have checked that these corrections are small at low $k$ due to the transfer function $T(k)$ in the real universe (though this would not be the case in pure Einstein de Sitter), and will therefore was not included in our numerics. On the other hand, the contribution from the UV part of $P_{13}$ contains the leading departure from SPT and is connected to the fluid parameters, as we have computed.

Related to this is how much the effective stress-tensor $[\tau^{ij}]_\Lambda$ varies from patch to patch.
This is measured by contractions of the variance
\bea
\mbox{Var}[\tau]^{ii'jj'}_\Lambda\!(t,t')\amp=\amp\langle[\tau^{ij}]_\Lambda(t)\,[\tau^{i'j'}]_\Lambda(t')\rangle
-\langle[\tau^{ij}]_\Lambda(t)\rangle\,\langle[\tau^{i'j'}]_\Lambda(t')\rangle
\eea
If we contract over $i,j$ and $i',j'$, and set $t'=t$, then this measures the fluctuations in the pressure; a type of stochastic pressure
\beq
(\Delta p)^2={1\over 9}\left(\langle[\tau]_\Lambda^2\rangle-\langle[\tau]_\Lambda\rangle^2\right)
\eeq
This is the fluctuations around the background value given by the mean $p_b={1\over 3}\langle[\tau]_\Lambda\rangle$ (\ref{bkgdpressure}). 
The absolute pressure does not affect the Newtonian dynamics (although it does affect the metric in GR). Instead the important fluctuations are those that arise from derivatives of the stress-tensor and are connected to the UV part of the $P_{22}(k)$ contribution to the power spectrum. These are small at low $k$, due to the transfer function in high $k$-modes as we mentioned previously, but can play an important role at higher $k$ approaching $k_{NL}$.

\end{document}